\documentclass[a4paper, 12pt]{article}
\usepackage{mathtext}               
\usepackage[T1]{fontenc}

\usepackage[utf8]{inputenc}         
\usepackage[english]{babel}
\usepackage{algorithm}
\usepackage{algpseudocode}
\usepackage{amsmath,amsfonts,amssymb,amsthm,mathtools}
\usepackage{icomma}
\usepackage{dsfont}
\usepackage{listings}
\usepackage{xcolor}
\usepackage{calc}

\usepackage{subcaption}
\usepackage{euscript}              
\usepackage{mathrsfs}              

\makeatletter
\renewcommand\Huge{\@setfontsize\Huge{14pt}{10}}
\renewcommand\huge{\@setfontsize\huge{14pt}{10}}
\renewcommand\Large{\@setfontsize\Large{14pt}{10}}
\renewcommand\large{\@setfontsize\large{12pt}{10}}
\makeatother

\usepackage[left=3cm,right=1.5cm,top=2cm,bottom=2cm,bindingoffset=0cm]{geometry}

\usepackage{enumitem}
\makeatletter
\AddEnumerateCounter{\asbuk}{\russian@alph}{щ}
\makeatother

\usepackage{caption}
\usepackage{graphicx}                  
\graphicspath{{images/}{images2/}}     
\usepackage{wrapfig}                   

\usepackage{array,tabularx,tabulary,booktabs} 
\usepackage{longtable}                        
\usepackage{multirow}                         
\usepackage{tikz}
\usetikzlibrary{trees, positioning, arrows.meta, decorations.pathreplacing, calc, graphs, graphs.standard}

\tikzset{
  treenode/.style = {
    circle,
    draw,
    rounded corners=4pt,
    align=center,
    minimum width=0.2cm,
    minimum height=.2cm,
    font=\sffamily,
    fill=blue!10
  },
  edge style/.style = {
    -{Stealth[length=2.5mm,width=1.8mm]},
    thick
  }
}

\usepackage{tikz}
\usetikzlibrary{graphs,graphs.standard,arrows.meta, shadows, positioning,fit,calc}

\usepackage{fancyhdr}
\usepackage{svg}
\svgpath{{./}} 

\usepackage{float}                   
\usepackage{gensymb}                 
\usepackage{listings}                
\usepackage[unicode, pdftex]{hyperref}

\usepackage{indentfirst}

\theoremstyle{remark}

\theoremstyle{definition}
\newtheorem{definition}{Definition}[section]

\newcommand{\beq}{\begin{equation}}
\newcommand{\eeq}{\end{equation}}
\newcommand{\beqnn}{\begin{equation*}}
\newcommand{\eeqnn}{\end{equation*}}

\newcommand{\diag}{\operatorname{diag}}

\newcommand{\sut}{\mathbf{SU}(2)}
\newcommand{\sun}{\mathbf{SU}(2^n)}
\newcommand{\un}{\mathbf{U}(2^n)}
\newcommand{\uf}{\mathbf{U}(4)}
\newcommand{\ut}{\mathbf{U}(2)}

\newcommand{\RR}{\mathbb{R}}

\newcommand{\Pg}{|0\rangle\langle 0|}
\newcommand{\Pu}{|1\rangle\langle 1|}
\newcommand{\ot}{\otimes}

\tikzset{
  modern/.style={
    rectangle,
    rounded corners=8pt,
    minimum width=3.5cm,
    minimum height=1.2cm,
    text centered,
    font=\sffamily\bfseries,
    fill=#1!15,
    draw=#1!80!black,
    drop shadow={shadow xshift=0.8ex, shadow yshift=-0.8ex, opacity=0.15}
  },
  arrowstyle/.style={
    -{Stealth[length=3mm,width=2mm]},
    thick,
    draw=black!85,
    rounded corners=8pt
  },
  notarrowstyle/.style={
    thick,
    draw=black!85,
    rounded corners=8pt
  }
}

\newcommand{\stackblock}[3][]{%
    \coordinate (temp) at (0,0);
    \ifx&#1&
        \coordinate (stackpos) at (0,0);
    \else
        \node[#1, opacity=0] (stackpos) {};
    \fi
    \node[modern=#2, opacity=0.6] (shadowtwo#3) at ([shift={(0.25,0.25)}]stackpos) {};
    \node[modern=#2, opacity=0.8] (shadowone#3) at ([shift={(0.12,0.12)}]stackpos) {};
    \node[modern=#2] (#3) at (stackpos) {#3};
}

\makeatletter
\newcounter{subsubsubsection}[subsubsection]

\renewcommand{\@tocrmarg}{2.55em plus1fil} 

\newcommand\subsubsubsection[1]{\@startsection{subsubsubsection}{4}{\z@}%
  {-3.25ex\@plus -1ex \@minus -.2ex}%
  {1.5ex \@plus .2ex}%
  {\normalfont\normalsize\bfseries}{#1}\ignorespaces} 

\newcommand{\l@subsubsubsection}{\@dottedtocline{4}{6.0em}{4.5em}}
\makeatother

\renewcommand\theparagraph{\thesubsubsection.\arabic{paragraph}}
\renewcommand\thesubparagraph{\theparagraph.\arabic{subparagraph}}

\usepackage{titlesec}
\titleformat{\paragraph}[block]{\normalsize\bfseries}{\theparagraph}{1em}{}
\titlespacing*{\paragraph}{0pt}{3.25ex plus 1ex minus .2ex}{1em}
\titleformat{\subparagraph}[block]{\normalsize\bfseries}{\thesubparagraph}{1em}{}
\titlespacing*{\subparagraph}{0pt}{3.25ex plus 1ex minus .2ex}{1em}

\begin{document}
    \begin{titlepage}

\vfill

\flushright{
ITEP/TH-18/26
\\
IITP/TH-16/26
\\
MIPT/TH-16/26
\\
}

\begin{center}
   \baselineskip=16pt
   {\large \bf Interpretable Machine Learning for Building Analytic Quantum Circuits: the Case of Diagonal Operators
   }
   \vskip 1cm
    Matvei Fedin$^{a,b,c}$\footnote{fedin.mm@iitp.ru, fedin.mm@phystech.edu},
    and Andrey Morozov$^{a,b,c,d}$\footnote{morozov.andrey.a@iitp.ru}
       \vskip .6cm
            \begin{small}
                     {\it
                          $^a$Moscow Institute of Physics and Technology, 141702, Dolgoprudny, Russia
                           \\
                          $^b$Institute for Information Transmission Problems, 127051, Moscow, Russia
                          \\
                          $^c$ITMO, 197101, St. Petersburg, Russia
                          \\
                          $^d$ITEP, Moscow, Russia
                          }
\end{small}
\end{center}

\vfill
\begin{center}
\textbf{Abstract}
\end{center}

\center{Machine Learning has become a useful instrument across many fields, but the opacity of typical ``black-box'' models limits their value for scientific discovery. In this paper we present a methodology that uses \emph{interpretable} Machine Learning to generate analytic mathematical hypotheses about a physical system, which can then be confirmed by a formal derivation. We demonstrate the methodology on the problem of decomposing unitary operators into elementary quantum gates. By training an interpretable model on circuits obtained numerically and analyzing its weights, we are led to a hypothesis for the analytic structure of the decomposition. Applied to the family of diagonal operators, this hypothesis yields a universal, analytic quantum circuit valid for an arbitrary number of qubits, whose $CNOT$ count scales as $\mathcal{O}(2^n)$, improving on the $\mathcal{O}(2.5^n)$ of standard numerical methods. The resulting decomposition is proved rigorously in a companion paper.}

\vfill
\setcounter{footnote}{0}
\end{titlepage}
    \newpage

    \tableofcontents{}
    \newpage

    \newcommand{\qnum}[1]{\left[#1\right]}
\newcommand{\brak}[1]{\left\langle #1 \right\rangle}
\newcommand{\A}{\mathcal{A}}
\newcommand{\SU}{\operatorname{SU}}

\section{Introduction}
\label{sec:Chapter0} \index{Chapter0}

Quantum computing is one of the promising answers to the growing complexity of computational problems, since it can offer an exponential speed-up for certain tasks. New hardware, however, brings new constraints and requires new approaches to building working programs. A program for a quantum computer is a unitary matrix, but not every unitary matrix can be written directly onto the hardware. As with classical computers, there is a fixed set of elementary operations, called gates, that the device is able to perform. Quantum programming therefore poses a new problem: how to expand a given matrix into a product of such elementary gates.

Frequently the problem under study depends on parameters, and one is interested in how the answer varies with them. This calls for an algorithm that depends on these parameters explicitly, so that its construction need not be repeated for every set of values; in other words, we want an analytic expression for the algorithm. A good example is the construction of knot polynomials \cite{Witten1989}; \cite{Reshetikhin1990}. Because of the connections between knot theory and topological quantum computing \cite{KITAEV20032}; \cite{Nayak2008}, there is a natural motivation to build such an algorithm, although doing so is far from straightforward. Here the answer is sought as a polynomial in a parameter $q$, so we need an analytic form of the algorithm that depends on $q$. The problem discussed in this paper originates in knot polynomials, but the result we obtain is considerably more general.

While Machine Learning (ML) is commonly used to build quantum circuits, this is usually done numerically and separately for each particular problem. In this paper we propose instead to use \emph{interpretable} ML to obtain an analytic form of the decomposition. The object we construct is a circuit decomposition, not a quantum algorithm in the usual sense (a sequence of steps that solves a computational task): the present work belongs to the field of quantum circuit compilation.

Recent progress in ML is increasingly accompanied by serious problems stemming from the lack of interpretability inherent in complex ``black-box'' models \cite{10127717}; \cite{BlackBoxModels}. Such models often hide the rationale behind their predictions \cite{zschech2025inherently}, which raises concerns about their transparency and reliability as decision-support systems \cite{Garouani_2024}. This limitation is especially critical for scientific discovery, since the inability to extract human-understandable relationships prevents the integration of ML results into the broader body of scientific knowledge \cite{rowan2025definitionimportanceinterpretabilityscientific}. Work on the interpretability of artificial intelligence often relies on methods of implicit interpretability \cite{joshi2025locallyparetooptimalinterpretationsblackbox}, which still do not explain the meaning of the individual weights of the trained model.

There is also a class of physics-informed ML models. These are still black boxes, but they are trained using information about physical laws \cite{watson2025machinelearningphysicsknowledge}; \cite{fink2025physicsmachinelearningback}. Incorporating such knowledge increases their reliability on physics problems compared with models trained on data alone \cite{watson2025machinelearningphysicsknowledge}; \cite{wu2025physicsinformedmachinelearningcombustion}, yielding more accurate models while reducing the amount of training data required \cite{nguyen2025physixfoundationmodelphysics}; \cite{wiesner2025physicsfoundationmodel}. They also help to shorten the simulation time of physical systems \cite{BALASHOV2025113554}, lower computing-power requirements, and simplify the scaling of numerical experiments.

To address this problem and contribute to genuine scientific understanding, our study deliberately uses interpretable ML models whose structure and parameters are based on the established physical principles governing the system under study. For example, in Section~\ref{sec:methodology} we choose a semi-linear model, motivated by the assumption that maps between the coordinates of linear spaces - such as the algebras of unitary groups and tensor products of these algebras - may themselves have linear subsystems or even be completely linear. We then analyze the weights of the trained model in order to build a theoretical description of the phenomena and to identify the relationships at work in our problem. Further considerations on interpretable ML can be found in \cite{duvenaud2013structurediscoverynonparametricregression}; \cite{WShITY}.

The centerpiece of our methodology is to formulate a mathematical hypothesis and then prove it rigorously, through a formal mathematical or physical derivation of every significant result first suggested by the analysis of the ML model. In this way we bridge the gap between data-driven research and fundamental theoretical understanding \cite{methodsinterpretableAI}; \cite{Craven1996Extracting}. This approach lets ML serve as a powerful tool for generating hypotheses within a framework that still requires final theoretical verification, so that the results obtained are more trustworthy and free from the pitfalls discussed in \cite{bioAI}.

The key idea is to use ML as a means of quickly forming a hypothesis without a deep prior analysis of the problem; once such a hypothesis is available, constructing a theoretical justification becomes much easier.

This paper is therefore devoted to a workflow for constructing a mathematical hypothesis with ML, starting only from the initial information about a system and its general properties. We demonstrate this workflow on the decomposition of quantum operators into elementary gates: training an interpretable model on numerically generated circuits and analyzing its weights reveals the analytic structure of the decomposition. For the family of diagonal operators this structure, derived in Section~\ref{sec:D_case}, constitutes our central hypothesis. It yields a universal, analytic quantum circuit valid for an arbitrary number of qubits, with a $CNOT$ count that scales as $\mathcal{O}(2^n)$ rather than the $\mathcal{O}(2.5^n)$ of standard numerical methods, and it is proved rigorously in the companion paper \cite{fedin2025mathematicalaspectsdecompositiondiagonal}.


As our physical problem we study quantum circuit compilation --- the decomposition of unitary operators into elementary gates. In plain terms, we try to decompose various subgroups and subsets of $\un$ operators into products of operators taken from a fixed basic set.

We consider only matrices from $\un$, because only unitary operations are physically possible on a quantum computer. Decomposing them into elementary operations is necessary in order to implement a given unitary operator on a quantum computer of a particular architecture.

Let $n$ be the number of qubits in the circuit. There are many quantum computer architectures, but in every case a universal quantum computer can efficiently realize the one-qubit operations of $\ut$. We also need an element of $\uf$ that entangles the states of two qubits; in our case this element is the $CNOT$ gate, described in more detail in Section~\ref{sec:intro_qc}.

Different decomposition methods can produce circuits of different \emph{depth} --- the maximum number of sequential gates acting on any single qubit. The depth should be minimized, both to improve the quality of the output and to reduce the runtime on the device.

By the algorithmic complexity of a circuit we mean how the depth required to model quantum operators of a given type grows with the number of qubits.

In the general case $U\in\un$, decomposing a matrix into a quantum circuit has complexity $\sim\mathcal{O}(n^2 4^n)$ \cite{nch}, so solving the problem in full generality is of little practical use.

\begin{figure}[h!]
        \centering
        \includegraphics[trim = 10 0 10 0, clip = true,width=.8\linewidth]{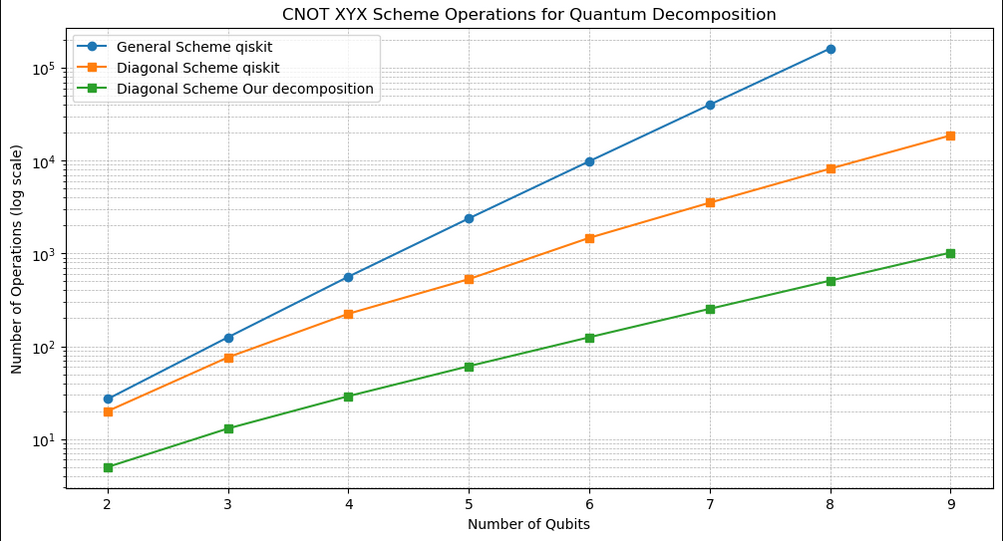}
        \caption{Number of elementary gates (logarithmic scale) in the $n$-qubit decomposition of a diagonal operator versus the number of qubits $n$, for the general \texttt{qiskit} decomposition, the \texttt{qiskit} diagonal decomposition, and our optimized decomposition. The values are those of Table~\ref{tab:cnot_xyx_decomposition}.}
        \label{fig:log_intro_plot}
\end{figure}

Our results mainly concern the diagonal operator, for which a much more complete form was obtained; see Table~\ref{tab:cnot_xyx_decomposition} and the logarithmic plot in Fig.~\ref{fig:log_intro_plot}.
The well-known numerical methods implemented in the \texttt{qiskit} library \cite{IBMQuantum2024} show an exponential dependence: in general their complexity is close to $\sim\mathcal{O}(n^2 4^n)$, while for diagonal operators it falls to $\sim\mathcal{O}(2.5^n)$.
The method we present here achieves $\sim\mathcal{O}(2^n)$ for diagonal matrices, which opens up clear opportunities for optimization.

\begin{table}[h!]
\centering
    \begin{tabular}{|c|c|c|c|}
        \hline
        \textbf{Qubits} & \textbf{G(\texttt{qiskit})} & \textbf{D(\texttt{qiskit})} & \textbf{D(optimized)}\\
        \hline
        2 & 27 & 20 & 5\\
        3 & 125 & 76 & 13\\
        4 & 562 & 224 &29 \\
        5 & 2387 & 529 &61\\
        6 & 9847 & 1472 &125\\
        7 & 40023 & 3520 &253\\
        8 & 161341 & 8192 &509\\
        9 & - & 18688 &1021\\
        \hline
        \end{tabular}
        \caption{Number of gates in the \(n\)-qubit decomposition into $CNOT$ and rotations, as a function of the number of qubits. \textbf{G(\texttt{qiskit})}: general \texttt{qiskit} decomposition; \textbf{D(\texttt{qiskit})}: \texttt{qiskit} decomposition of a diagonal operator; \textbf{D(optimized)}: our optimized decomposition of a diagonal operator.}
        \label{tab:cnot_xyx_decomposition}
        
\end{table}

\newpage

\subsection{Software and hardware}\label{sec:setup}

All numerical decompositions in this work are obtained with the \texttt{qiskit} library~\cite{IBMQuantum2024}, used solely as a numerical source of circuit decompositions; the optimization of the interpretable model is carried out with \texttt{pytorch}~\cite{paszke2019pytorch}. The choice of these particular libraries is incidental: the methodology operates on the resulting parameter vectors and does not depend on them. All experiments were run on the following machine:

\begin{table}[h!]
\centering
\begin{tabular}{ll}
\toprule
Component & Specification \\
\midrule
OS  & Debian 12 (Bookworm) \\
CPU & AMD Ryzen 9 (16 cores) \\
RAM & 128 GB \\
GPU & NVIDIA GeForce RTX 4060 Ti (16 GB) \\
\bottomrule
\end{tabular}
\caption{Hardware used for the experiments.}
\label{tab:hardware}
\end{table}

\newpage

\section{Introduction into Machine Learning}
\label{sec:ml_methods_ref}

Before the main text, it is  worth mentioning basic definitions and concepts from Machine Learning, some of which will be used later in this article.

Machine Learning (ML) is a class of artificial intelligence methods aimed at building mathematical models capable of revealing hidden patterns in empirical data. Unlike traditional programming, where the algorithm is specified explicitly, in ML, the model selects its internal parameters in a process called learning to solve the problem. The process of building a model includes several fundamental components.

\subsection{Data}
The basis for training is a data set or sample \(\mathcal{D} = \{(x_i, y_i)\}_{i=1}^N\), where \(x_i\in \mathbb{R}^d\)~--- is the feature vector of a \(d\)-dimensional object, and \(y_i\)~--- is the corresponding value of the target variable. Depending on the type of \(y_i\), the tasks are divided into regression tasks (if \(y_i\in\mathbb{R}^k\)~--- vector of continuous quantities) and classification problems (if \(y_i\)~--- is a discrete class label).
For an objective assessment of the quality and generalizing ability of the model, as well as to prevent overfitting~--- situations where the model perfectly describes the training data, but loses the ability to predict new data, --- the initial sample is divided into three disjoint subsets:
\begin{itemize}
    \item {Training set:} is used for direct optimization of model parameters.
    \item {Validation set:} is used to select {hyperparameters} (architectural parameters that cannot be adjusted during training, for example, the degree of a regression polynomial or the learning rate).
    \item {Test set:} Is used for the final, independent evaluation of the performance of the final model on data that did not participate in the training.
\end{itemize}

\subsection{Model and Architecture}

Model \(f(x; \Theta)\) is a parametric family of functions that map the feature space to the target variable space. Here \(\Theta\)~ is a vector of configurable parameters (weights) of the model. For example, for a linear model the mapping looks like:

\begin{equation}
\hat{y} = f(x; W, b) = Wx + b, 
\label{eq:main_linear_model}
\end{equation}

where \(\hat{y}\)~--- model prediction, \(W\)~--- the weight matrix, \(b\)~--- the shift vector. The interpretability of such a model is maximal, since the weights of the matrix ($W$) directly characterize the linear relationship between the input and output variables, and are convenient for analyzing and searching for the fundamental causes of this dependence. 

The architecture of the model \(\mathcal{A}\) is its formal mathematical template that defines the class of functions that the model is able to approximate, but does not depend on the specific dimensions of the problem. In turn, parameters such as the dimension of the input (\(M\)) and output (\(N\)). We refer to the hyperparameters that define a specific representative of the architecture.
  
Thus, we can introduce a mapping that, for a given architectural template \(\mathcal{A}\) and hyperparameters \((M,N)\), generates a specific parametric family of models (hypotheses) \(\mathcal{H}_{\mathcal{A}}(M,N)\).

\begin{equation}
\mathcal{A}: (\mathbb{N}, \mathbb{N}) \rightarrow \mathcal{H}_{\mathcal{A}}(M, N)    
\end{equation}

A specific model \(f_{\Theta}\) having trainable parameters \(\Theta\) is an element of this space: \(f_{\Theta}\in\mathcal{H}_{\mathcal{A}}(M,N)\). For example, if the linear regression architecture is selected as \(\mathcal{A}\), then for the mapping task \(\mathbb{R}^M\rightarrow\mathbb{R}^N\) it generates a hypothesis space \(\mathcal{H}_{\text{lin.reg.}}(M,N)= \{ f(x)= Wx+b\mid W\in\mathbb{R}^{N\times M}, b\in\mathbb{R}^N \}\).

The key methodological principle of this work is to fix the architecture \(\mathcal{A}\) at all stages of the study. However, when the conditions of the problem change (for example, when analyzing systems with different numbers of qubits), the hyperparameters \(M\) and \(N\) will change. Therefore, we will use the same conceptual architecture, but in each case we will deal with different representatives of it, adapted to the appropriate dimension of the task.

\subsection{Loss function}
To quantify the discrepancy between the prediction of the model \(\hat{y}_i = f(x_i; \Theta)\) and the true value of \(y_i\) we use Loss Function \(\mathcal{L}(\hat{y}_i, y_i)\). In regression problems, the classic choice is Mean Squared Error (MSE):
\begin{equation}
    \mathcal{L}(\hat{y}_i, y_i) = |y_i - \hat{y}_i|^2.
\end{equation}
The learning process (optimization of Loss Function) consists in finding a set of parameters \(\Theta^*\) that minimizes empirical risk --- the average loss function over the entire training sample:
\begin{equation}
\Theta^* = \arg\min_{\Theta} \frac{1}{N} \sum_{i=1}^{N} \mathcal{L}(f(x_i; \Theta), y_i).
\end{equation}

Minimization is usually carried out by iterative numerical methods, for example, gradient descent method, where the parameters are updated at each step in the direction of the empirical risk anti-gradient.

\subsection{Metrics}

For a final assessment of the generalizing ability and practical applicability of the model, quality metrics are used, which are calculated on a test sample. Unlike the loss function, the metric does not have to be continuous and differentiable; its choice is dictated solely by the application requirements of the problem \cite{Terven_2025}. For regression problems, common metrics are:
\begin{itemize}
    \item {Mean Absolute Error, (MAE):} \( \text{MAE} = \frac{1}{N} \sum_{i=1}^{N} |y_i - \hat{y}_i| \).
    \item {Root Mean Squared Error (RMSE):} \( \text{RMSE} = \sqrt{\frac{1}{N} \sum_{i=1}^{N} |y_i - \hat{y}_i|^2} \).
    \item {Coefficient of determination (\(R^2\)):}\(R^2 =1 - \frac{\sum_{i=1}^{N}|y_i - \hat{y}_i|^2}{\sum_{i=1}^{N} |y_i - \bar{y}|^2}\), where \(\bar{y} = \frac{1}{N}\sum_{i=1}^N{y_i}\) is the average value of true responses in the sample. A value close to $1$ indicates the high quality of the model.
\end{itemize}

\subsection{Padding}

In many ML tasks, models are designed to process input data in fixed-size batches to leverage the parallel computing capabilities of modern hardware (such as GPUs). However, real-world data often comes in sequences of different lengths. For instance, in quantum computing case, some scheme in set can have different numbers of quantum gates. To combine these disparate examples into a single tensor for batch processing, all sequences must be brought to a uniform length.

Padding is the process of adding special dummy values, usually zeros, to shorter sequences so that they match the length of the longest sequence in the batch (or a predefined maximum length). Consider a batch of three sequences with lengths $3$, $5$, and $4$. After padding to a maximum length of $5$, they become:

\[
\begin{array}{ll}
\text{Original:} & [w_1, w_2, w_3] \\
\text{Padded:}    & [w_1, w_2, w_3, 0, 0] \\[4pt]
\text{Original:} & [w_1, w_2, w_3, w_4, w_5] \\
\text{Padded:}    & [w_1, w_2, w_3, w_4, w_5] \\[4pt]
\text{Original:} & [w_1, w_2, w_3, w_4] \\
\text{Padded:}    & [w_1, w_2, w_3, w_4, 0]
\end{array}
\]

Here, \(w_i\) represents a feature vector for an element of the sequence, and \(0\) denotes the padding token.

\subsection{Regularization}

As noted above, one of the central problems of ML is overfitting, in which the model, having an excessively large number of free parameters, adjusts not to the general patterns of the data, but to random noise inherent in a particular training sample. A characteristic symptom of this phenomenon is the uncontrolled growth of the absolute values of the weights \(\Theta\), which makes the model excessively sensitive to small fluctuations in the input features. Regularization is a set of methods aimed at limiting the effective complexity of the model and improving its generalizing ability.

The most common approach is to modify the optimized functional by adding a penalty term (regularizer) \(\Omega(\Theta)\), which penalizes undesirable configurations of the weights. The empirical risk minimization problem then takes the form:
\begin{equation}
\Theta^* = \arg\min_{\Theta} \left[ \frac{1}{N} \sum_{i=1}^{N} \mathcal{L}(f(x_i; \Theta), y_i) + \lambda\, \Omega(\Theta) \right],
\end{equation}
where \(\lambda \geq 0\)~--- is the regularization coefficient, a hyperparameter that controls the balance between the accuracy of describing the training data and the smoothness (simplicity) of the model. At \(\lambda = 0\) the problem degenerates into ordinary minimization of empirical risk, and with an increase in \(\lambda\) the requirement for the smallness of the weights is strengthened. Depending on the type of the penalty term \(\Omega(\Theta)\), two classical types of regularization are distinguished:
\begin{itemize}
    \item {\(L_2\)-regularization (weight decay):} the penalty term is defined as the square of the Euclidean norm of the weight vector, \(\Omega(\Theta) = \|\Theta\|_2^2 = \sum_{j} \theta_j^2\). This approach evenly reduces the absolute values of all weights, preventing individual parameters from becoming disproportionately large, and leads to smoother mappings.
    \item {\(L_1\)-regularization:} the penalty term is set as the norm \(L_1\), \(\Omega(\Theta) = \|\Theta\|_1 = \sum_{j} |\theta_j|\). Its distinctive property is the tendency to zero out insignificant weights, which leads to sparse solutions and can be used for the implicit selection of the most informative features.
\end{itemize}

Along with the modification of the loss function, there are structural methods of regularization specific to neural networks. Among them, the dropout method, which consists in randomly disabling (zeroing) a certain proportion of neurons at each iteration of training, is widely used. This prevents the co-adaptation of neurons and forces the network to form more robust and redundant internal representations. Like the coefficient \(\lambda\), the fraction of the disabled neurons is a hyperparameter selected on the validation set.

    \section{Introduction into Quantum Computing \label{sec:intro_qc}}

\subsection{Decomposition in Quantum Computing context}

In quantum computing, decomposition refers to the representation of an arbitrary unitary operator as a sequence of elementary operations from a fixed set of basis gates. This procedure has three fundamental aspects, each of which is important for the practical implementation of quantum algorithms.

The first aspect concerns hardware limitations. Modern quantum processors can only perform a limited set of elementary operations, such as one- and two-qubit gates. This means that an arbitrary unitary operator must be decomposed into a sequence of possible elementary operations.

The second aspect concerns the optimization of quantum circuits. For practical implementation on a real quantum computer, the decomposition process must minimize the number of operations --- this is critical to reducing the impact of decoherence and other noise effects. The shorter the sequence of operations, the higher the probability of obtaining a correct result. The different precisions of one- and two-qubit operations on a real quantum computer need special attention: according to the study \cite{expcnotsign}, typical gate fidelities are $99.7\%$ and $96.5\%$ for one- and two-qubit gates, respectively. The error probability therefore differs by roughly an order of magnitude, which forces us to minimize the number of two-qubit operations in our circuits.

The third aspect concerns standardization. Decomposition into universal bases, such as the set $\{CNOT, H, T\}$, implies the universality of quantum algorithms across different hardware platforms. Moreover, in the case of a finite number of basic gates, we can only speak of a decomposition with a given precision, as is evident from the cardinality of the sets: the group of unitary matrices is continuous, and the group generated by a finite number of gates is at most countable. If we use $\{|0\rangle, |1\rangle \}$ as a basis, we can write basic operators as:

\begin{equation}
    |0\rangle = \begin{bmatrix}
        1\\0
    \end{bmatrix};     |1\rangle = \begin{bmatrix}
        0\\1
    \end{bmatrix}; 
    H = \frac{1}{\sqrt{2}}\begin{bmatrix}
        1&1\\1&-1
    \end{bmatrix}; 
    T = \begin{bmatrix}
        1&0\\0&e^{i\pi/4}
    \end{bmatrix}; 
\end{equation}

\begin{equation}
     X = \begin{bmatrix}
        0&1\\1&0
    \end{bmatrix}; CNOT = \Pg \ot I + \Pu \ot X
\end{equation}

This is especially important given the existence of different quantum processor architectures. In other words, if we know how to simulate the gates $\{CNOT, H, T\}$ on a given quantum processor, then we know how to execute any quantum algorithm on this quantum processor with a given accuracy, according to the Solovay-Kitaev theorem \cite{kitaev1997}; \cite{kitaev2002}.

However, we can approach this issue differently: many quantum computer architectures allow us to perform, generally speaking, any one-qubit operation. Therefore, in addition to $CNOT$, we can include any $\mathbf{U}(2)$ matrices in our basic operations. This approach allows us to express any operators exactly.

Any unitary operator $U \in \mathbf{U}(2)$ can be represented as:

\begin{equation}
U = e^{i\alpha} R_x(\beta) R_y(\gamma) R_x(\delta)
\label{eq:u_standart_def}
\end{equation}
where $R_y$ and $R_x$ correspond to rotations around the $Y$ and $X$ axes, respectively. This representation is known as a rotation sequence decomposition.

For multi-qubit systems, such as operators from $U \in \mathbf{SU}(4)$, more complex decomposition methods are used. Among these, cascade decompositions such as the Cartan decomposition are particularly useful. 

The Cartan (KAK) decomposition of a two-qubit unitary operator $U$ is a factorization that represents an arbitrary two-qubit gate as a sequence of three special blocks:

\begin{equation}
U = (U_1 \otimes U_2) \cdot \exp (i (X\ot X h_x + Y\ot Y h_y + Z\ot Z h_z) ) \cdot (U_3 \otimes U_4), \text{ where, } U_i \in \mathbf{U}(2)
\label{eq:weylcam}
\end{equation}

Geometrically, the parameters $h_x$, $h_y$ and $h_z$ in~\eqref{eq:weylcam} are coordinates in the Weyl chamber --- a prism \cite{crooks2024} that provides a canonical representation of the set of all such operators and in which only the \textit{interaction} part is taken into account.

A special case is represented by diagonal operators, for which, as we will see later in Section \ref{sec:comp_exp}, only the $h_z$ parameter in the Weyl chamber is non-zero. This property allows the operator to be implemented using only two $CNOT$ gates. This result has important practical implications for optimizing quantum circuits.

\subsection{The quantum gates universality}

A set of quantum gates is called \emph{universal} if an arbitrary unitary operator on $n$ qubits can be \emph{approximated} with any given precision $\epsilon$ by a finite sequence of gates from this set. In contrast to the usual meaning, there is an approximation here, because having a finite number of matrices, multiplying them will not be possible to obtain a continuum group, so it is precisely the possibility of dense coverage of the continuum group \cite{nch}, \cite{kitaev1997}, \cite{kitaev2002} that is used.

An example of a universal set in the sense of quantum computing is $\{CNOT,\,H,\,T\}$\cite{nch}. Whereas if we replace $\{H,T\}$ with $\sut$, we get an exact universal set of $\{CNOT,\, \sut\}$, which is universal in the usual sense.

The Solovay–Kitaev theorem \cite{kitaev1997}; \cite{kitaev2002} guarantees that if we can form a dense subgroup everywhere, then approximating any operator in it with precision $\epsilon$ is effective: the length of the gate chain grows as a polynomial of $\log(1/\epsilon)$.

Quantum entanglement is a phenomenon in which the state of a group of particles cannot be represented as a tensor product of the states of individual particles, even if the particles are located at a great distance from each other.

The Clifford group $\mathcal{C}_n$ for $n$ qubits is defined as the normalizer of the Pauli group $\mathcal{P}_n$ in the unitary group:
\begin{equation}     \mathcal{C}_n = \left\{ U \in U(2^n) \mid U\,\mathcal{P}_n\,U^\dagger = \mathcal{P}_n \right\} \end{equation} 

Quantum Circuits consisting only of Clifford gates are effectively simulated classically in accordance with the Gottsman–Knill theorem~\cite{nch}.

The addition of the gate $T = \mathrm{diag}(1,\,e^{i\pi/4})$ (also known as the $\pi/8$–gate) to the Clifford valves makes the set $\{\mathrm{Clifford},\,T\}$ universal in an approximate sense: any unitary operator can be reconstructed with arbitrary precision using a finite sequence of these gates.

\subsection{Parametrization of diagonal operators}

Because this paper focuses on diagonal operators, we record here the parametrization of the diagonal subgroup that the rest of the paper relies on.

Let us consider the problem of parameterization of an arbitrary unitary operator in quantum computing. The basic idea is based on the fundamental relationship between the parameters of the diagonal operator and the parameters of the quantum circuit. For a group of $\mathbf{SU}(N)$ special unitary matrices, where $N=2^n$ for a system of $n$ qubits, the Cartan decomposition $U =K_1 D K_2$ plays a key role, where $K_1, K_2$ belong to the Clifford subgroup, and $D$ is a diagonal matrix of Cartan's subgroups.

Mathematically speaking, diagonal matrices form a maximal commutative subalgebra (Cartan subalgebra) $\mathfrak{h}$ in the Lie algebra $\mathfrak{su}(N)$. The basis of this subalgebra can be chosen as follows. Consider a set of matrices $H_k\in\mathfrak{su}(N)$, $k=1,\ldots,N-1$, where each $H_k$ has the form:

\begin{equation}    
H_k = \diag(
\underbrace{0,\ldots,0}_{k-1\text{ times}}
,1,-1,
\underbrace{0,\ldots,0}_{N-k-1\text{ times}}
) 
\end{equation}
from $1$ to $k$-th position and $-1$ to $(k+1)$-th position. These matrices are linearly independent and commute between each other $[H_i,H_j] = 0$ for all $i,j$, which makes them a natural choice for the Cartesian subalgebra basis.

Any diagonal matrix $D\in\mathbf{SU}(N)$ can be parameterized using an exponential map:
\begin{equation}     D(\theta_1,\ldots,\theta_{N-1}) = \exp\left(i\sum_{k=1}^{N-1} \theta_k H_k\right) = \diag\left(e^{i\lambda_1},\ldots,e^{i\lambda_N}\right) 
\label{eq:D_old_decomp}\end{equation}
where $\theta_k\in\RR$ are real parameters, and the phases of $\lambda_j$ are expressed in terms of $\theta_k$ in a linear manner. It is important to note that from the condition of the unit determinant $\sum_{j=1}^N\lambda_j=0\mod 2\pi$, which guarantees matrix $D$ belongs to $\mathbf{SU}(N)$.

The decomposition \eqref{eq:D_old_decomp} is equivalent to our decomposition, which is more convenient for practical tasks:

\begin{equation}     D(\theta_1,\ldots,\theta_{N-1}) = \exp\left(i\sum_{k=1}^{N-1} \theta_k M_k\right) = \diag\left(e^{i\theta_1},e^{i\theta_2},\ldots,e^{-i(\theta_1+\ldots+\theta_{N-1})}\right) \in \mathbf{SU}(N) \end{equation}

Where we use matrices $M_k$ with $1$ to the $k$-th position and $-1$ on the last position for the same range of $k$:

\begin{equation}    
M_k = \diag(\underbrace{0,\ldots,0}_{k-1\text{ times}},
1,\underbrace{0,\ldots,0}_{N-k-1\text{ times}},
-1) 
\end{equation}

Since these coordinates transform into each other linearly, if there is a linear mapping between the circuit parameters and the Cartan decomposition, then the mapping between the quantum circuit parameters and the diagonal operator parameters will be linear in our basis.
If we also want to consider matrices from $\mathbf{U}(N)$ then we just need to replace the last angle with a free one, instead of the sum of the remaining angles with a minus:

\begin{equation}
    D(\theta_1,\ldots,\theta_{N}) = \diag\left(e^{i\theta_1},e^{i\theta_2},\ldots,e^{i\theta_{N}}\right) \in \mathbf{U}(N) 
    \label{eq:D_U}
\end{equation}

    
\newpage 
    \section{Methodology \label{sec:methodology}}
\label{sec:Chapter3} \index{Chapter3}

Our method for finding a simple parametric mapping in the decomposition of quantum circuits is based on analyzing the parameters of ML models trained on the parameters of a large number of quantum circuits. These circuits are obtained from a numerical decomposition routine; we use the numerical decomposition routine of Section~\ref{sec:setup}, but this is just the software implementation we chose. The methodology operates on the resulting parameter vectors alone and is therefore portable: any routine that decomposes a unitary into circuit parameters can be used without changing the method. This lets us use modern computational methods to reduce manual analysis, although some stages still require direct human involvement. A more detailed description of the workflow is given in Section~\ref{sec:wfg}. We first motivate the use of ML methods for analyzing quantum circuits.

\subsection{Motivation of the methodology}

We want to obtain possible factorizations of matrices from a given $\un$ subgroup as quantum circuits using an interpretable ML model. We first want to use a semi-linear model: model that is the sum of a linear model and a three-layer fully-connected model. We add regularization to the loss function based on the weights of the fully connected model to bias the model toward a linear response.
In the example from Section ~\ref{sec:comp_exp}, we will see that, at least for diagonal matrices of dimension $4$, there is a linear mapping between the circuit parameters and the matrix parameters.

\subsection{Workflow graph\label{sec:wfg}}

In this section, we will take a detailed look at how our method works in the form of a convenient graphical representation with detailed explanations of each step. We will consider a generalization of the previous considerations and consider in what format we can study classes of subgroups. As we have already noted, on a quantum computer with $n$ qubits, we can simulate only operators from the $\sun$ group. It is worth noting that from the point of view of quantum computing, the groups $\sun$ and $\un$ are indistinguishable, so we will choose the most appropriate in each specific case. 

In the following parts, we will focus on studying workflow to formulate a mathematical hypothesis for a certain subgroup of $G\leq\un$ using ML methods.

\subsubsection{Main graph \label{sec:main_graph}}

\begin{figure}[h!]
    \centering
        \scalebox{0.75}{

    \begin{tikzpicture}[node distance=1.0cm]
    
      \stackblock{blue}{Raw Data};
      
      \node (q) [modern=teal, right=.5cm of Raw Data] {QC scheme};
      \node (a) [modern=green, below=of Raw Data] {model};
      \node (u) [modern=orange, below= of a] {MWA};
      \stackblock[below=6.5cm of Raw Data]{blue}{Pretty Data};

      \node (s) [modern=teal, right= of Pretty Data] {simplified QC scheme};

      \node (a1) [modern=green, below=of Pretty Data] {model};
      \node (u1) [modern=orange, below= of a1] {MWA};
      \node (r) [modern=red, below= of u1] {Math Hypothesis};
      \draw[arrowstyle] (Raw Data) -- (a);
      \draw[arrowstyle] (a) -- (u);
      \draw[arrowstyle] (s) -- (Pretty Data);
      \draw[arrowstyle] (Pretty Data) -- (a1);
      \draw[arrowstyle] (a1) -- (u1);
      \draw[arrowstyle] (u1) -- (r);

      \draw[arrowstyle] (q) |- (u);
      \draw[arrowstyle] (s) |- (u1);
      \node (group1) [draw=black!70, dashed, rounded corners,
                fit=(q) (a) (u),
                inner sep=12pt, label=left:{Stage $1$}] {};
        
        \draw[arrowstyle] 
            (u) -- ++(0,-1.35)
            -- ++(1,0)
             -| (s);

      \node (group2) [draw=black!70, dashed, rounded corners,
                fit=(s) (a1) (u1),
                inner sep=12pt, label=left:{Stage $2$}] {};
                
    \end{tikzpicture}    
    }
    \caption{The two-stage workflow for forming a mathematical hypothesis. Stage~1 produces a simplified circuit from Raw Data; Stage~2 produces the hypothesis from Pretty Data. MWA denotes model-weight analysis.}
    \label{fig:workflow-graph}
\end{figure}

A general view of the data processing and hypothesis formulation scheme in two stages is shown in Fig. \ref{fig:workflow-graph}.
In general, it can be seen that the hypothesis development scheme works in two main stages: Stage $1$ and Stage $2$. In Stage $1$, we are more interested in blind research: we try to obtain Raw Data from the structure of the subgroup $G\leq\un$ using standard mathematical techniques together with the numerical decomposition routine. For more information about how to get data from the group structure, see the Section \ref{sec:data-preparation}. Here we will focus more on the general features of the algorithm.

Let's say we got Raw Data linking the scheme parameters and the parameters of the elements of the $G$ group in a suitable parameterization, as well as the general form of the circuit, as returned by the decomposition routine, for which this decomposition holds. The problem with this scheme in particular and such data in general is that the number of parameters in the scheme and in the group may be different. This means that numerical methods cannot obtain the minimum and optimal expansion, as we have already noted in the introduction.
However, we assume that these data can still be adequately described by the model.

On the other hand, now the mapping from the parameters of quantum circuits to the parameters of a group element is surjective: we have several configurations of the parameters of quantum circuits that are mapped into a single group element. From the point of view of the ML task, this means that we will not be able to train a deterministic model to accurately predict the parameters of a quantum circuit based on the parameters of a group element, since it will always, at best, bring one possible element out of the set and unambiguously minimize the convergence of the loss function. In this case, a good solution is to try to find a mapping from the parameters of the quantum circuit to the parameters of the group elements. ML models are good at understanding that multiple input data elements are projected into the same element, but not vice versa \cite{Forets_2023};\cite{liu2020deepneuralarchitectureslosing};\cite{jiang2025surjectivityneuralnetworkselicit}.

Based on physical, mathematical and other considerations, for example, considerations on the topic of believing in the simplicity of the answer, we must choose the architecture of the model that will be used on both Stages, possibly using different representatives of it. The reasons for using different architecture representatives in this case are due to the fact that after optimizing the quantum circuit, the number of quantum operations and, consequently, the number of parameters may decrease.

Let's train the model on Raw Data. In this case, we will get some matrix of weights of the model, the dimensions of which is equal to $N$ and $M$ respectively.

The next step, MWA (model weight analysis), involves analyzing the model's weights (parameters). This can involve calculating statistical moments of the model's parameters or performing a visual inspection if the model's size allows. This analysis should take into account the context of the model's weights, the order of the arguments, and the quantum circuit. It is clear that if the repetitive patterns present in the QC scheme are repeated in the weights of the model, then this indicates that they are of the same nature and can be excluded from the scheme, reducing its redundancy. Some combinations of quantum gates are also possible, giving multiplication by a representative of a group with constant parameters. In this case, it introduces some bias weights into the ML model, and when we remove this section of the quantum chain, we can remove bias, thereby optimizing the model.

Based on the MWA, we can hypothesize a minimal quantum scheme, the simplified QC scheme. This is no longer just one of the possible decomposition schemes obtained numerically, but a scheme claiming to be a general decomposition. We then move to Stage~$2$. At this stage we again use the decomposition routine to generate many instances of the simplified QC scheme; the details of this process are described in Section~\ref{sec:data-preparation}.

Now, following the diagram in Fig.~\ref{fig:workflow-graph}, we perform the same steps, except that we swap the inputs and the outputs. The aim at this stage is a model that is easy to invert and whose weights admit a clear interpretation, so that the MWA step can synthesize a mathematical hypothesis from them. In favorable cases the weight matrix itself has a regular algebraic structure - for example a block or recursive pattern - and recognizing this structure is exactly what turns the trained model into the hypothesis, which is then proved separately.

\subsubsection{Data Preparation \label{sec:data-preparation}}

In this section we look more closely at how the data are prepared for our ML model at the two stages. Schematically, the construction of Raw Data for Stage~$1$ is shown in Fig.~\ref{fig:Raw-Data}, and that of Pretty Data for Stage~$2$ in Fig.~\ref{fig:Pretty-Data}. We examine each in turn.

\paragraph{Raw Data.} First of all, we fix the subgroup $G\leq\un$ that we want to explore. Its elements are parametrized as described in Section~\ref{sec:intro_qc}: through the exponential map over a chosen basis of the corresponding subalgebra (for the diagonal case, Eq.~\eqref{eq:D_old_decomp}). As argued there, the coordinates transform into each other linearly, so we expect the map between the operator parameters and the circuit parameters to be natural, and possibly linear or quadratic. Parametrizing $M$ representatives of $G$ in this way yields the parameter vectors $\{\vec y_1,\vec y_2,\dots,\vec y_M\}\subset\RR^l$; for a diagonal operator, $\vec y_i$ collects the angles $\theta_k$ of Eq.~\eqref{eq:D_old_decomp}. Decomposing the same elements $\{g_1,g_2,\dots,g_M\}\subset G$ into quantum circuits with the decomposition routine gives the circuit-parameter vectors $\{\vec x_1,\vec x_2,\dots,\vec x_M\}\subset\RR^k$. Now we should look at the problems we will encounter along the way.

\begin{figure}[h!]
    \centering
    \scalebox{0.75}{

    \begin{tikzpicture}[node distance=1.0cm]
          
      \node (G) [modern=red] {$G\leq \un$};
      \node (g) [modern=yellow, below = of G] {$\{g_1,g_2,\dots,g_M\}\subset G$};
      \node (P) [modern=red, left = of g] {Parametrization};
      \node (y) [modern=yellow, below = of P] {$\{\vec y_1,\vec y_2,\dots,\vec y_M\}\subset \RR^l$};
      \node (F) [modern=green, below = of y] {Filterring};
      \node (D) [modern=green, right = of g] {Decompose};
      \node (x) [modern=yellow, below = of D] {$\{\vec x_1,\vec x_2,\dots,\vec x_M\}\subset \RR^k$};
      \node (PCA) [modern=green, below = of x] {Clasterization};
      \node (Y) [modern=yellow,below = of F] {$Y = \{\vec y_{i_1},\vec y_{i_2},\dots,\vec y_{i_m}\}$};
      \node (X) at (G |- Y)[modern=yellow] {$X = \{\vec x_{i_1},\vec x_{i_2},\dots,\vec x_{i_m}\}$};

      \node (group1) [draw=black!70, dashed, rounded corners,
                fit=(X) (Y),
                inner sep=12pt, label=left:{Dataset}] {};
      \stackblock[below = of group1]{blue}{Raw Data};
      \node (qs) at (PCA |- Raw Data)[modern=teal] {QC scheme};

      \draw[arrowstyle] (G) -- (g);
      \draw[arrowstyle] (g) -- (P);
      \draw[arrowstyle] (g) -- (D);
      \draw[arrowstyle] (D) -- (x);
      \draw[arrowstyle] (x) -- (PCA);
      \draw[arrowstyle] (PCA) -- (qs);
      \draw[arrowstyle] (P) -- (y);
      \draw[arrowstyle] (y) -- (F);

\draw[arrowstyle] (PCA) -- (F);

      \draw[arrowstyle] (PCA) -| (X);
      \draw[arrowstyle] (F) -- (Y);
      \draw[arrowstyle] (group1) -- (Raw Data);

    \end{tikzpicture}    
    }
    \caption{Generation of Raw Data for Stage~1: group elements are parametrized and, in parallel, decomposed into circuits; the circuits are clustered and both are filtered into a single dataset.}
    \label{fig:Raw-Data}
\end{figure}

First of all, as mentioned before, the decomposition is not unique. It is also obvious that the mappings between the parameters of the group and the parameters of the quantum circuit in the decomposition strongly depend on the specific type of quantum circuit, which means that we must somehow filter out unnecessary circuits. There are many ways to do this, of varying complexity and optimality. One suboptimal and difficult-to-implement approach is to attempt to analyze the form of each quantum circuit individually as a circuit object. In this case, one would have to consider the sequence of quantum operators by their types (e.g., $R_X$, $CNOT$, etc.) and then consider their parameters separately. This is a dead end, because due to software limitations, this is only possible on a processor with a high latency per object. The most optimal and widely used alternative in ML is the PCA method, which is presented in more details in Appendix~\ref{sec:PCA}, followed by clustering. Geometrically, the parameters \(\{\vec{x}_1, \vec{x}_2, \dots, \vec{x}_M\}\) form several clouds of points in parameter space---clusters. To illustrate, consider galaxies in the universe: a galaxy is a cluster of stars, and a cluster here is a cluster of points. Each such cluster corresponds to a specific configuration of quantum circuit. We should also note that different configurations need not have the same dimension. This means that we zero-pad all vectors to the dimensionality of the highest-dimensional vector, allowing them to be clustered in a unified space.

Anyway, the clustering step provides us with a quantum scheme that corresponds to a selected cluster suitable for analysis and a list of data indexes corresponding to this class. At the Filtering step, we leave those elements of the group whose indexes are contained in this list using Hierarchical Clusterization, since the same object corresponds to the same index in one case as a parametrization of a quantum operator, and in the second as a parametrization of a quantum circuit. Since dumping into a certain decomposition cluster is a rather stochastic process, we can assume that with such a selection we get the same coverage of the group $G$ that is quite dense in the $\epsilon$ grid (in the sense that in each $\epsilon$ cell, there is almost certainly at least $1$ in the set of acceptable parameters representative (with probability tending to $1$)). In the future, we will verify this statement after the fact.

This ready-made Raw Data is used in Stage $1$ as described in Section \ref{sec:main_graph}.

We show an example of using this scheme:

\begin{enumerate}
    \item Let's choose $G$ as a diagonal subgroup of $\un$ for $n=2$.
    \item Let's use $\Big\lbrace \operatorname{diag}\left(e^{\frac{i}{5}}, e^{\frac{i\sqrt{2}}{5}}, e^{\frac{i\sqrt{3}}{5}}, e^{\frac{i\sqrt{4}}{5}}\right),$ 
    $ \operatorname{diag}\left(e^{\frac{i\sqrt{5}}{5}}, e^{\frac{i\sqrt{6}}{5}}, e^{\frac{i\sqrt{7}}{5}}, e^{\frac{i\sqrt{8}}{5}}\right), $
    
    $\operatorname{diag}\left(e^{\frac{i\sqrt{9}}{5}}, e^{\frac{i\sqrt{10}}{5}}, e^{\frac{i\sqrt{11}}{5}}, e^{\frac{i\sqrt{12}}{5}}\right)$ $\Big\rbrace$ as a subset of $G$.
    \item Let's use parametrization $\diag\left(e^{i\alpha_1}, e^{i\alpha_2}, e^{i\alpha_3}, e^{i\alpha_4}\right)$. Now we have map  $\RR^4 \xrightarrow{} G$.
    \item Now we have $\vec y_1 = \begin{bmatrix}
        \frac{\sqrt{1}}{5}\\\frac{\sqrt{2}}{5}\\\frac{\sqrt{3}}{5}\\\frac{\sqrt{4}}{5}\\
    \end{bmatrix}$; $\vec y_2 = \begin{bmatrix}
        \frac{\sqrt{5}}{5}\\\frac{\sqrt{6}}{5}\\\frac{\sqrt{7}}{5}\\\frac{\sqrt{8}}{5}\\
    \end{bmatrix}$;
    $\vec y_3 = \begin{bmatrix}
        \frac{\sqrt{9}}{5}\\\frac{\sqrt{10}}{5}\\\frac{\sqrt{11}}{5}\\\frac{\sqrt{12}}{5}\\
    \end{bmatrix}$
    \item Filtering this data for different types of QC (look at QC example, Fig. \ref{fig:gate_numbers}) requires decomposing a set of unitary operators with the decomposition routine. This set consists of $\{\vec x_1, \vec x_2, \vec x_3\}$ where \newline $\vec x_1 = (
\frac{\pi}{2} \  \frac{\pi}{2} \ -\frac{\pi}{2} \ -\frac{\pi}{2} \ \frac{\pi}{2} \ -\frac{\pi}{2} \ \frac{\pi}{2} \ -\frac{\pi}{2} \ 0.0146 \ 0.0 \ \frac{\pi}{2} \ 0.0682 \ \frac{\pi}{2} \ 1.439 \ -\frac{\pi}{2})$
\newline $\vec x_2 = (
\frac{\pi}{2} \  \frac{\pi}{2} \ -\frac{\pi}{2} \ -\frac{\pi}{2} \ \frac{\pi}{2} \ -\frac{\pi}{2} \ \frac{\pi}{2} \ -\frac{\pi}{2} \ 0.0031 \ 0.0 \ \frac{\pi}{2} \ 0.0396 \ \frac{\pi}{2} \ 1.492 \ -\frac{\pi}{2})$
\newline $\vec x_3 = (
-\frac{\pi}{2} \  \frac{\pi}{2} \ -\frac{\pi}{2} \ \frac{\pi}{2} \ \frac{\pi}{2} \ -\frac{\pi}{2} \ \frac{\pi}{2} \ -\frac{\pi}{2} \ 0.0015 \ 0.0 \ -\frac{\pi}{2} \ 0.0309 \ \frac{\pi}{2} \ 1.633 \ \frac{\pi}{2})$
\item Using clusterization we can get two clusters: $C_1$ and $C_2$. $C_1 = \{\vec x_1,\  \vec x_2\}$ and $C_2 = \{\vec x_3\}$ with $|C_1| = 2$ and $|C_2| = 1$. $|C_1|>|C_2|$, so we choose $C_1$.
\item Filter set of $y$ by indices from set $C_1$ we get set of two elements: $\{\vec y_1, \ \vec y_2\}$
\item Now we have $X = \{\vec x_1,\  \vec x_2\} $ and $Y = \{\vec y_1,\  \vec y_2\} $. It's Raw Data. We also should leave only the changing variables for $X$:
\begin{equation}
    X = \left\lbrace\begin{bmatrix}
        0.0146\\0.0682\\-0.131\\
    \end{bmatrix}, \begin{bmatrix}
        0.0031\\0.0396\\-0.078\\
    \end{bmatrix}\right\rbrace
\end{equation}

\end{enumerate}

\paragraph{Pretty Data. \label{sec:gen_pretty_date}} The Stage $1$ returns us a simplified QC scheme, using which we should generate Pretty Data. The main difference between Pretty Data and Raw Data is that we are aware not only of the structure of the $G$ subgroup, but also of the structure of the general decomposition of this subgroup.  In this case, we assume that we will get all the elements of $G$ from all QCs of a given shape. This sounds plausible, because the surjection of quantum circuits is just supposed to compress everything to the $G$ group.

Anyway, now we are not generating the parameters of the group $Y$, but the parameters of the quantum circuit $X$, making up a simplified QC scheme with these parameters, considering its unitary operator and parametrizing it as $Y$. Now it is important that the same quantum operator can be represented by its parameters in different ways due to symmetry over a period multiple of $\pi$. To avoid this behavior, we generate circuit parameters from a sufficiently small neighborhood of zero $U_{\epsilon}(0)$ for some $\epsilon$. In this case, the newly obtained data should be clustered again and the largest or most suitable cluster should be selected for analysis, filtered out $X$ as in the previous stage and combined into Pretty Data. This is the end of the data generation stage.

\begin{figure}[h!]
    \centering
    \scalebox{0.75}{
    \begin{tikzpicture}[node distance=1.0cm]

      \node (x) [modern=yellow] {$\{\vec x_1,\vec x_2,\dots,\vec x_M\}\subset U_{\epsilon}(0)$};
      \node (qs) at (PCA |- Raw Data)[modern=teal, below = of x] {simplified QC scheme};
      \node (g) [modern=yellow, below = of qs] {$\{g_1,g_2,\dots,g_M\}\subset G$};
      \node (P) [modern=red, left = of g] {Parametrization};
      \node (y) [modern=yellow, below = of P] {$\{\vec y_1,\vec y_2,\dots,\vec y_M\}\subset \RR^l$};
      \node (PCA) [modern=green, right = of y] {Clasterization};
      \node (F) [modern=green, right = of PCA] {Filterring};

      \node (Y) [modern=yellow,below = of PCA] {$Y = \{\vec y_{i_1},\vec y_{i_2},\dots,\vec y_{i_m}\}$};
      \node (X) [modern=yellow,below = of F] {$X = \{\vec x_{i_1},\vec x_{i_2},\dots,\vec x_{i_m}\}$};

      \node (group1) [draw=black!70, dashed, rounded corners,
                fit=(X) (Y),
                inner sep=12pt, label=left:{Dataset}] {};
      \stackblock[below = of group1]{blue}{Pretty Data};

      \draw[arrowstyle] (x) -- (qs);
      \draw[arrowstyle] (qs) -- (g);
      \draw[arrowstyle] (g) -- (P);
      \draw[arrowstyle] (P) -- (y);
      \draw[arrowstyle] (y) -- (PCA);
      \draw[arrowstyle] (PCA) -- (Y);
      \draw[arrowstyle] (PCA) -- (F);
      \draw[arrowstyle] (F) -- (X);
      \draw[arrowstyle] (group1) -- (Pretty Data);
      \draw[arrowstyle] (x) -| (F);

    \end{tikzpicture}    
        }
    \caption{Generation of Pretty Data for Stage~2: circuits are generated from the simplified scheme, their operators parametrized, then clustered and filtered into a single dataset.}
    \label{fig:Pretty-Data}
\end{figure}

We show an example of using this scheme:

\begin{enumerate}
    \item Let's choose neighborhood $X\subset U_{\epsilon}(0)$
    \item Now, using QC scheme 
    \item Let's use parametrization $\diag\left(e^{i\alpha_1}, e^{i\alpha_2}, e^{i\alpha_3}, e^{i\alpha_4}\right)$. Now we have map  $\RR^4 \xrightarrow{} G$.
    \item Now we have $\vec y_1 = \begin{bmatrix}
        \frac{\sqrt{1}}{5}\\\frac{\sqrt{2}}{5}\\\frac{\sqrt{3}}{5}\\\frac{\sqrt{4}}{5}\\
    \end{bmatrix}$; $\vec y_2 = \begin{bmatrix}
        \frac{\sqrt{5}}{5}\\\frac{\sqrt{6}}{5}\\\frac{\sqrt{7}}{5}\\\frac{\sqrt{8}}{5}\\
    \end{bmatrix}$;
    $\vec y_3 = \begin{bmatrix}
        \frac{\sqrt{9}}{5}\\\frac{\sqrt{10}}{5}\\\frac{\sqrt{11}}{5}\\\frac{\sqrt{12}}{5}\\
    \end{bmatrix}$
    \item Filtering this data for different types of QC requires decomposing a set of unitary operators with the decomposition routine. This set consists of $\{\vec x_1, \vec x_2, \vec x_3\}$ where \newline $\vec x_1 = (
\frac{\pi}{2} \  \frac{\pi}{2} \ -\frac{\pi}{2} \ -\frac{\pi}{2} \ \frac{\pi}{2} \ -\frac{\pi}{2} \ \frac{\pi}{2} \ -\frac{\pi}{2} \ 0.0146 \ 0.0 \ \frac{\pi}{2} \ 0.0682 \ \frac{\pi}{2} \ 1.439 \ -\frac{\pi}{2})$
\newline $\vec x_2 = (
\frac{\pi}{2} \  \frac{\pi}{2} \ -\frac{\pi}{2} \ -\frac{\pi}{2} \ \frac{\pi}{2} \ -\frac{\pi}{2} \ \frac{\pi}{2} \ -\frac{\pi}{2} \ 0.0031 \ 0.0 \ \frac{\pi}{2} \ 0.0396 \ \frac{\pi}{2} \ 1.492 \ -\frac{\pi}{2})$
\newline $\vec x_3 = (
-\frac{\pi}{2} \  \frac{\pi}{2} \ -\frac{\pi}{2} \ \frac{\pi}{2} \ \frac{\pi}{2} \ -\frac{\pi}{2} \ \frac{\pi}{2} \ -\frac{\pi}{2} \ 0.0015 \ 0.0 \ -\frac{\pi}{2} \ 0.0309 \ \frac{\pi}{2} \ 1.633 \ \frac{\pi}{2})$
\item Using clustering we obtain two clusters, $C_1$ and $C_2$. $C_1 = \{\vec x_1,\  \vec x_2\}$ and $C_2 = \{\vec x_3\}$ with $|C_1| = 2$ and $|C_2| = 1$. $|C_1|>|C_2|$, so we choose $C_1$.
\item Filter set of $y$ by indices from set $C_1$ we get set of two elements: $\{\vec y_1, \ \vec y_2\}$
\item Now we have $X = \{\vec x_1,\  \vec x_2\} $ and $Y = \{\vec y_1,\  \vec y_2\} $. It's Pretty Data.

\end{enumerate}

\subsubsection{Model variants}

All the methods for studying the decomposition of matrix subgroups that we discussed in previous sections work model-architecture-free. Because we can work normally and interpret a finite class of models: linear models and linear metamodels of any order or similar, although in the general case everything is limited by the imagination of the researcher.

In general, with such a workflow, the choice of model is left to the researcher and must be mathematically and physically motivated. In the example discussed in Section \ref{sec:Chapter4}, we use a semi-linear model. 

\subsection{Worked example: the two-qubit diagonal case}

Having described the workflow in the abstract, we now run it end-to-end on the simplest non-trivial case - a two-qubit diagonal operator - as an illustration. This example also produces the minimal two-qubit circuit that we reuse in Section~\ref{sec:Chapter4}.

\subsubsection{A single-parameter test case}

To verify the linearity of our mapping, we can look at the mappings between all the parameters of the diagonal matrix and the parameters of the quantum circuit. For clarity, it is worth finding a parametrization in which all the parameters of the quantum circuit and all the parameters of the unitary operator depend linearly on some common parameter. Here is an example from topological quantum computing (TQC)\cite{KITAEV20032}, in which, according to the article \cite{kolganov2022largektopologicalquantum}, the following diagonal operator occurs:

\begin{equation}
    \bar{T}^2(\varphi) = \diag(\exp(i\varphi(m+N-1)2m)\text{ for $m\in\{1,2,\dots, N\}$}) \in \mathbf{U}(N)
\label{eq:barTsq}
\end{equation}

Referring to the formula \eqref{eq:D_U}, it has a relation between the parameters:

\begin{equation}
    \theta_{m} = (m+N-1)2m\varphi, \quad \forall m \in \{1,2,\dots, N\}
    \label{eq:theta_VARPHI}
\end{equation}

It has positive coefficients in front of each parameter, since $(m+N-1)2m>0$ for all valid $m$ and $N$, and also depends on only one parameter. 
With this approach, if the parameters of the quantum circuit turn out to be linearly dependent on $\varphi$, this gives us good reason to believe that the mapping is linear (or semi-linear) in the general case, although it does not formally prove it.

We will also understand the form in which it is worth looking for an answer for the general case if we find one for our particular case.

Moreover, the two-qubit case is convenient because it has an upper limit on the number of $CNOT$s for different values of $h_i$ in the Weyl chamber.
\cite{threeCNOT}. See Appendix \ref{sec:threeCNOT} for more information.

\subsubsection{The two-qubit numerical experiment\label{sec:comp_exp}}

\begin{figure}[ht]
        \centering
        \includegraphics[trim=100 0 0 500,clip=true,width=.98\linewidth]{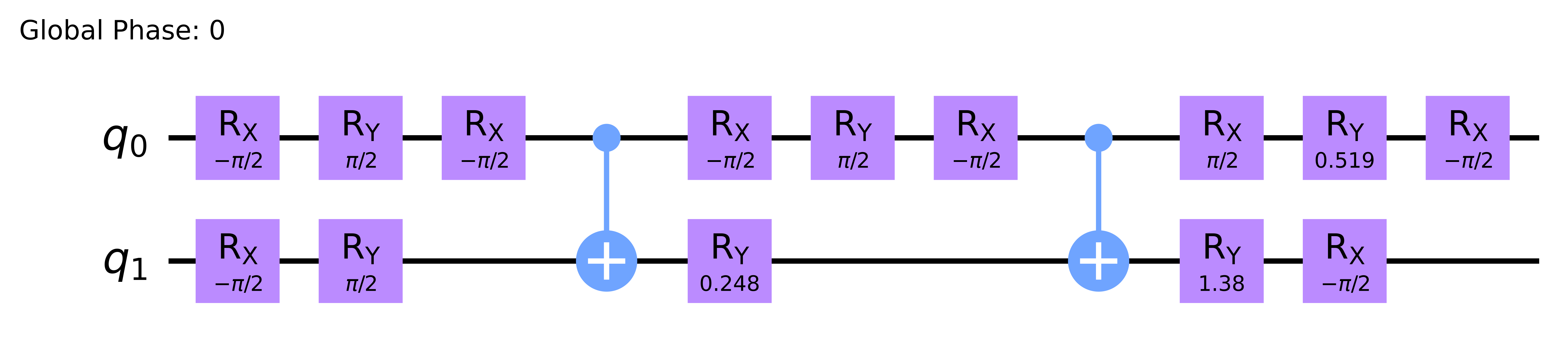}
        $\vec\theta_I = (0.82088788\,\ 0.05371843\,\ 0.76408384\,\ 0.49326067)^T$
        \centering
        \includegraphics[trim=100 0 0 500,clip=true,width=.98\linewidth]{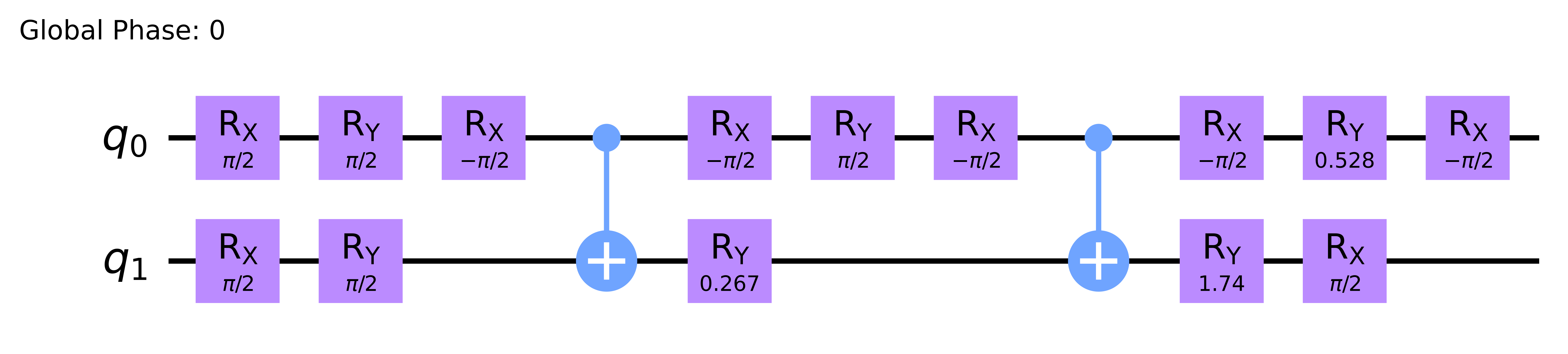}
        $\vec\theta_{II} = (0.83977211\,\ 0.04521547\,\ 0.74583697\,\ 0.48479206)^T$
        \caption{Comparison of two schemes having approximately similar final operators, but having very different parameters. $||\vec\theta_I - \vec\theta_{II}||_{\infty}<0.05$, but the difference in circuit parameters $\vec\beta$ is $||\vec\beta_{I} - \vec\beta_{II}||_{\infty} = \pi$}
        \label{fig:nearsch}
\end{figure}

In the first numerical experiment, we verify the existence of a linear map between the quantum circuit parameters and the diagonal operator parameters in the 2-qubit case. This result motivates our choice of the semi-linear architecture in the general case. To do this, using the decomposition routine, we simply get several decompositions for various $\varphi$ for a two-qubit scheme. It is clear that a numerical algorithm can find only one solution out of several possible ones. In practice, this means that diagonal operators with similar parameters: $|\vec\theta_{II}-\vec\theta_{I}|<\epsilon$ can have schemes with very different parameters, as in Fig. \ref{fig:nearsch}. We will look at the reasoning for this in more detail below. 

The reasons for such parameter jumps are the symmetries of the circuit, e.g.,

\begin{multline}
(X\ot I)\cdot CNOT\cdot (X\ot X) =\\= (X\ot I)\cdot (\Pg\ot I + \Pu\ot X)\cdot  (X\ot X) =\\= X\Pg X\ot X + X \Pu X \ot X^2 =\\= \Pu\ot X+\Pg \ot I = CNOT
\end{multline}

\begin{multline}
(I\ot R_X(\theta))\cdot CNOT\cdot (I\ot R_X(-\theta)) =\\= (I\ot R_X(\theta))\cdot (\Pg\ot I + \Pu\ot X)\cdot (I\ot R_X(-\theta)) =\\= (\Pg\ot (R_X(\theta)\cdot I\cdot R_X(-\theta)) + \Pu\ot (R_X(\theta)\cdot X\cdot R_X(-\theta))) =\\= \Pg\ot I+\Pu \ot X = CNOT
\end{multline}

\begin{figure}
    \centering
    \begin{minipage}[c]{0.3\linewidth}
        \centering
        \scalebox{.6}{\includegraphics[width=\linewidth]{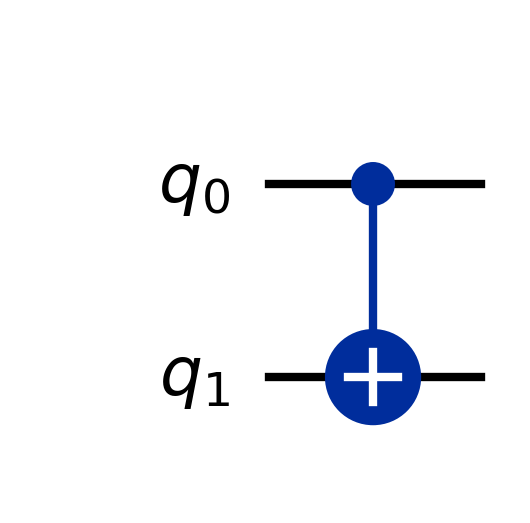}}
    \end{minipage}
    \begin{minipage}[c]{0.02\linewidth}
        \centering
        \Huge$=$
    \end{minipage}
    \begin{minipage}[c]{0.3\linewidth}
        \centering
        \scalebox{.99}{\includegraphics[width=\linewidth]{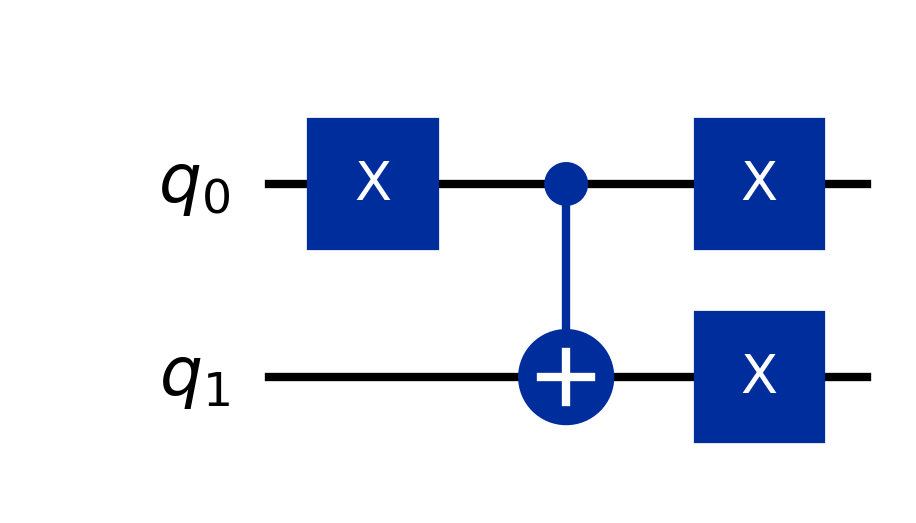}}
    \end{minipage}
    \hspace{0.25cm}
    \begin{minipage}[c]{0.02\linewidth}
        \centering
        \Huge$=$
    \end{minipage}
    \begin{minipage}[c]{0.3\linewidth}
        \centering
        \scalebox{.99}{\includegraphics[width=\linewidth]{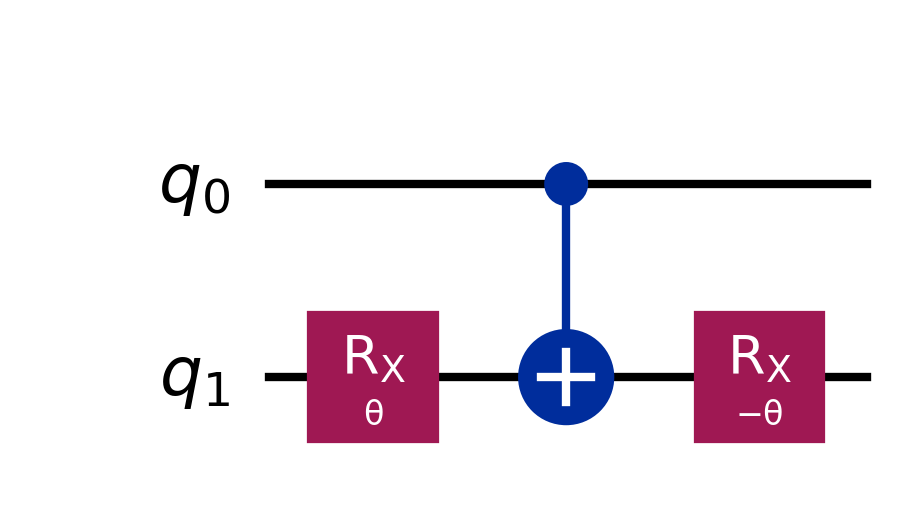}}
    \end{minipage}
    \caption{A CNOT gate and two circuits equivalent to it, used to explain the parameter jumps in the decomposition.}
    \label{fig:cnot}
\end{figure}

See also Fig. \ref{fig:cnot}. The actions of the operator $X$ represent the inversion of the ground states of the qubit (excited and ground) and its matrix in the computational basis is written as:
\begin{equation}
    X = \begin{pmatrix}
    0&1\\1&0
\end{pmatrix}
\end{equation}

In addition to the XYX type decomposition, there is, for example, a ZYZ type decomposition, which uses a sequence of operators $R_ZR_YR_Z$ by analogy with $R_XR_YR_X$ for the XYX decomposition. Based on a numerical experiment on the decomposition of diagonal matrices using the ZYZ decomposition type, it can be seen that the parameters of the linear mapping almost never converge, while XYX converges.
This is primarily due to the fact that, up to multiplication by $X$ on the left and/or right and the global phase, $XYX$ naturally converges to $R_Z$ ($R_X(-\frac{\pi}{2})R_Y(\theta)R_X(\frac{\pi}{2})  =R_Z(\theta)$), while $ZYZ$ converges to $R_Z(\beta_1)R_Y(0)R_Z(\beta_2)=R_Z(\beta_1+\beta_2)$, where the values of the parameters $\beta_1$ and $\beta_2$ are ambiguous, which adds a physically unobservable freedom to the system and the ambiguity in the decomposition.

\begin{figure}
    \centering
    \includegraphics[width=1\linewidth]{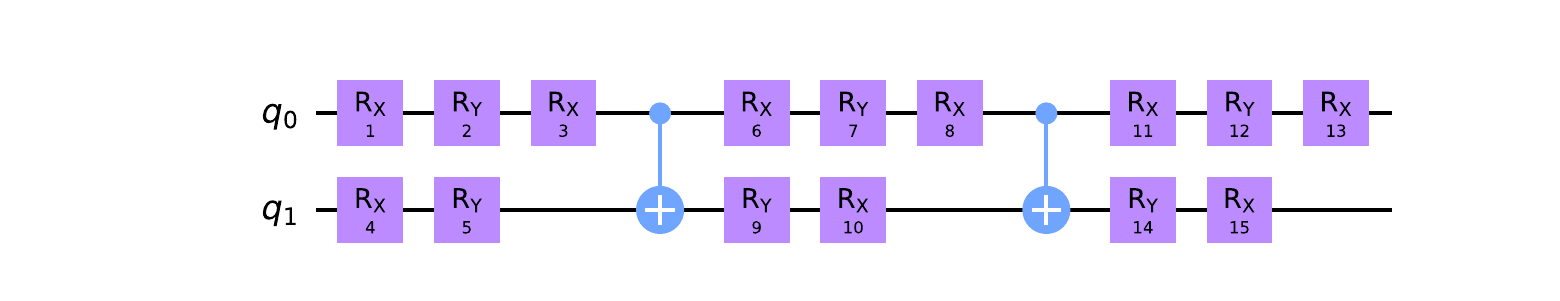}
    \caption{We represent the qubit parameters as a vector in $\RR^{15}$. The parameter numbers in the QC correspond to the indices in the vector.}
    \label{fig:gate_numbers}
\end{figure}

\begin{figure}
    \centering
    \includegraphics[width=\linewidth]{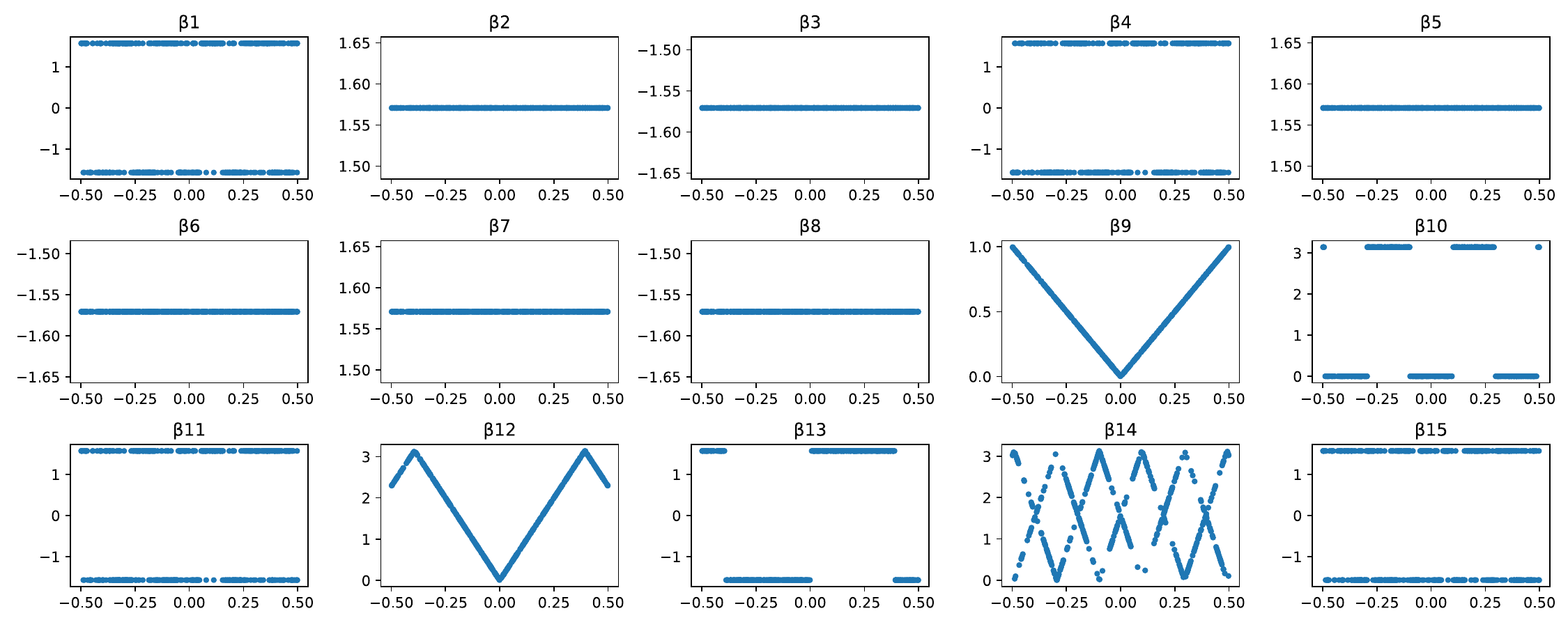}
    \caption{Plots of decomposition parameters $\bar{T}^2(\varphi)$, which show a set of points corresponding to a huge number of decompositions obtained numerically. These plots clearly show the linearity of the circuit parameters in accordance with $\varphi$ and the jumps explained using Fig. \ref{fig:cnot}}
    \label{fig:15_thetas_plot}
\end{figure}

\
\begin{figure}[h]
    \centering
    \includegraphics[width=.5\linewidth]{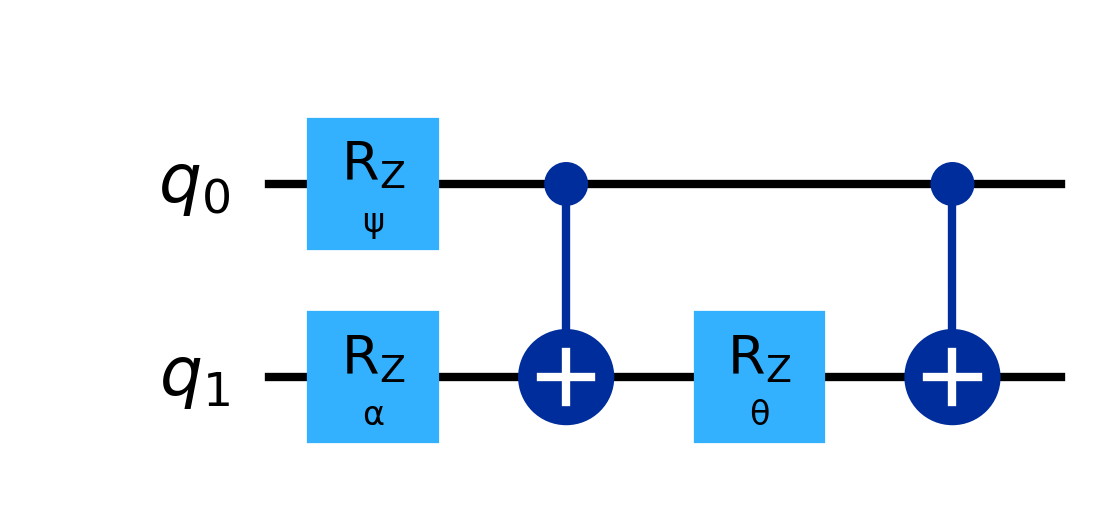}
    \caption{The simplified two-qubit diagonal scheme, with $2$ $CNOT$s and $3$ $R_Z$ gates.}
    \label{fig:simple_2qubit}
\end{figure}

As a result of numerical experiments on the decomposition of the matrix $\bar{T}^2(\varphi)$ for various $\varphi$, we obtained graphs of various parameters of the quantum circuit depending on $\varphi$ in Fig. \ref{fig:gate_numbers}, the number of the parameter responsible for the specified quantum gate is shown in Fig. \ref{fig:15_thetas_plot} we clearly see that $\beta_1, \beta_4, \beta_{11}, \beta_{15}$ have two possible values, the difference between which is $\pi$, which corresponds to the application of the operator $X$.

Now, using the properties in Fig. \ref{fig:cnot}, it is possible to reduce all triples of $XYX$ operators to the form $R_X(-\frac{\pi}{2})R_Y(\theta)R_X(\frac{\pi}{2}) =R_Z(\theta)$ and get a more beautiful scheme in Fig. \ref{fig:simple_2qubit}. The coefficients in front of the parameters are no longer important at this stage, because soon we will calculate them in general form.

Now we understand the two-qubit scheme and its minimal form through $2$ $CNOT$ and $3$ $R_Z$. 
    \newpage
\section{Experiments}
\label{sec:Chapter4} \index{Chapter4}

We deliberately choose the simplest interpretable model - a semi-linear model. As we can see below, for diagonal operators, the correction vanishes and the map turns out to be exactly linear is an empirical finding of the experiments below, recovered from the trained weights and proved separately in~\cite{fedin2025mathematicalaspectsdecompositiondiagonal}, rather than a built-in assumption. This exact linearity is what makes the weights directly readable: the recovered weight matrix $r_n$ of the already fully optimized scheme (Section~\ref{sec:tpf}) is itself the hypothesis.

Initial experiments confirmed our assumptions about the linearity of the mapping between $XYX$ and the standard two-qubit decomposition routine after all simplifications.

\begin{figure}[h!]
    \centering
    \includegraphics[width=\linewidth]{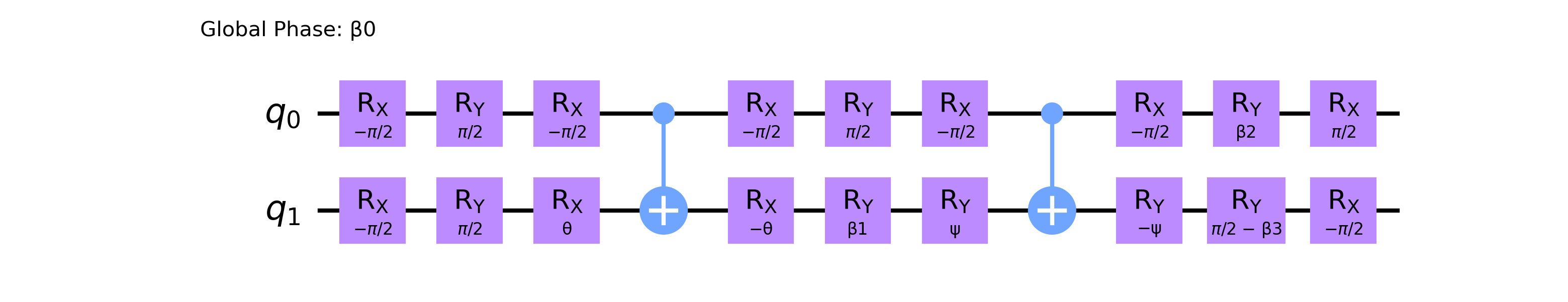}
    \caption{One possible design for a two-qubit diagonal operator.}
    \label{fig:2q_general_dec}
\end{figure}

We are now able, by examining a large number of generated decompositions, to understand their general form and free parameters (in the figure \ref{fig:2q_general_dec}, these are $\{\beta_i\}$, and $\psi$, $\theta$, parameters that do not affect the final answer, according to Fig. \ref{fig:cnot}).

We want to find a mapping from the parameters of the diagonal matrix $Y_{data}$ to the parameters of the quantum circuit $X_{data}$. We will do this by analyzing the generated data and training a model. As will be demonstrated in the experiment in Section \ref{sec:datagen2}, it is more optimal to generate the parameters of the diagonal matrix from the quantum circuit in Fig. \ref{fig:2q_general_dec} than to obtain the decomposition of a given diagonal matrix. Numerical decomposition is a computationally very complex process, while composition is simply a multiplication (possibly tensor multiplication) of unitary matrices. Furthermore, decomposition is generally not unique, as we already saw in Section \ref{sec:comp_exp}. This creates additional problems, as the model attempts to predict an average decomposition scheme, that is equally unsuitable for both configurations, but formally minimizes the mean square error. Data filtering at this stage is discussed in more detail in Section \ref{sec:datagen1}.

The logic in this section is based on the following:

\begin{enumerate}
    \item We want to obtain a general form of the circuit.
    \item We want to verify the correctness of our form of the circuit using computational methods.
\end{enumerate}

Since we're testing mappings of a certain type by training the corresponding ML model, we'll use the same model architecture both times, but the training and data preparation processes will differ.

In Section \ref{sec:datagen1}, we'll discuss how to assemble a dataset without problems, identified in Section \ref{sec:comp_exp} using only target diagonal matrices. In Section \ref{sec:tp}, we'll discuss the general subtleties of model training and the conclusions that can be drawn by training it on the datasets obtained in Section \ref{sec:datagen1}. In Section \ref{sec:datagen2}, we'll generate much larger datasets than in Section \ref{sec:datagen1}, knowing the response form in advance. In Section \ref{sec:tpf}, we'll verify the correctness of our assumption: the scheme we obtained in Section \ref{sec:tp} indeed produces only matrices from the set under consideration.

A proof that our scheme yields all matrices in the set under consideration will not be given here. We have only examined the methodology and the fundamental possibility of obtaining them. In the case of a diagonal matrix, we provided a proof in the article \cite{fedin2025mathematicalaspectsdecompositiondiagonal}

The full \texttt{Python} code with detailed comments is available in our \texttt{github} \cite{diagonal_decomposition} repository.

\subsection{Diagonal matrix expansion \label{sec:D_case}}

\subsubsection{Stage \texorpdfstring{$1$}{1} \label{sec:linear_model}}

This is Stage~$1$ of the workflow of Section~\ref{sec:wfg} (Fig.~\ref{fig:workflow-graph}), applied to the diagonal subgroup: we generate Raw Data, train the model, and read off a simplified circuit from the model weights at the MWA step.

The first experiment will be conducted on diagonal matrices as a subgroup of $\un$. As a model, we use the interpretable semi-linear model shown in Algorithm~\ref{lst:linear_model}.

\begin{algorithm}
\caption{Semi-linear model}\label{lst:linear_model}
\begin{algorithmic}[1]
\Require input dimension $d_{\text{in}}$, output dimension $d_{\text{out}}$
\State initialize a weight matrix $W\in\mathbb{R}^{d_{\text{out}}\times d_{\text{in}}}$ with entries drawn uniformly from $[-1,1]$, and a bias $b\in\mathbb{R}^{d_{\text{out}}}$
\State initialize a nonlinear component $g_{\phi}:\mathbb{R}^{d_{\text{in}}}\rightarrow\mathbb{R}^{d_{\text{out}}}$ with trainable parameters $\phi$
\Function{Model}{$x$}
    \State \Return $Wx + b + g_{\phi}(x)$
\EndFunction
\end{algorithmic}
\end{algorithm}

Its initialization type and device fully comply with the conditions that will be specified in detail in Section~\ref{sec:tp}, and the model will be trained by stochastic gradient descent (SGD) in Section~\ref{sec:tp}.

\paragraph{Dataset\label{sec:datagen1}}

In this section, we will take a closer look at the key stage of our research. We use the decomposition routine to study the formal decomposition of various diagonal quantum operators. The Section~\ref{sec:comp_exp}  described in detail the reasons for the surjectivity of the mapping we are looking for: there are many \textit{optimal} (that is, having the same number of quantum operators from a given set) quantum circuits for each diagonal unitary operator. There is an obvious problem that we will encounter when using the generated data directly - we will look for a mapping between the \textit{average} quantum circuit, given that it may not have a specific operator type. Strictly speaking, let's say we have two circuit configurations with parameter vectors $\vec{\theta_1}$ and $\vec{\theta_2}$ simulating a single unitary operator with a parameter vector $\vec{\varphi}$.
Then if we look for a map of $L:\theta\rightarrow\varphi$ by gradient descent to minimize the RMSE, then at each step we will get an approximation of $L_n$ for which:

\begin{equation}
    \lim_{n\rightarrow\infty}||L_n(\vec{\theta_1})- \vec{\varphi}||_2 = \lim_{n\rightarrow\infty}||L_n(\vec{\theta_2})- \vec{\varphi}||_2 \neq 0
\end{equation}

This is a rather serious problem, because, as indicated in the Section~\ref{sec:comp_exp}, there can be major changes in the scheme even with a small operator change.

Unlike the reverse mapping, a direct mapping from a quantum circuit to a quantum unitary operator gives a single, unambiguous result. In this case, the data obtained should be segmented into several dense clusters of points in the parameter space. 

At this stage, we cannot begin to slightly change the parameters of quantum circuits in order to obtain similar diagonal quantum operators, since many circuit parameters in the two-qubit case generally have a discrete set of values in Fig. \ref{fig:15_thetas_plot}, beyond which the result no longer will be a diagonal matrices.

We use the PCA (principal component analysis, described in more detail in Section ~\ref{sec:PCA}) algorithm to initially compress the data into a lower-dimensional space for further clustering into different quantum circuit configurations.

\begin{figure}[htbp]
    \centering
    \begin{minipage}{0.48\linewidth}
        \centering
        \includegraphics[ trim=00 0 00 0,  
    clip=true,width=\linewidth]{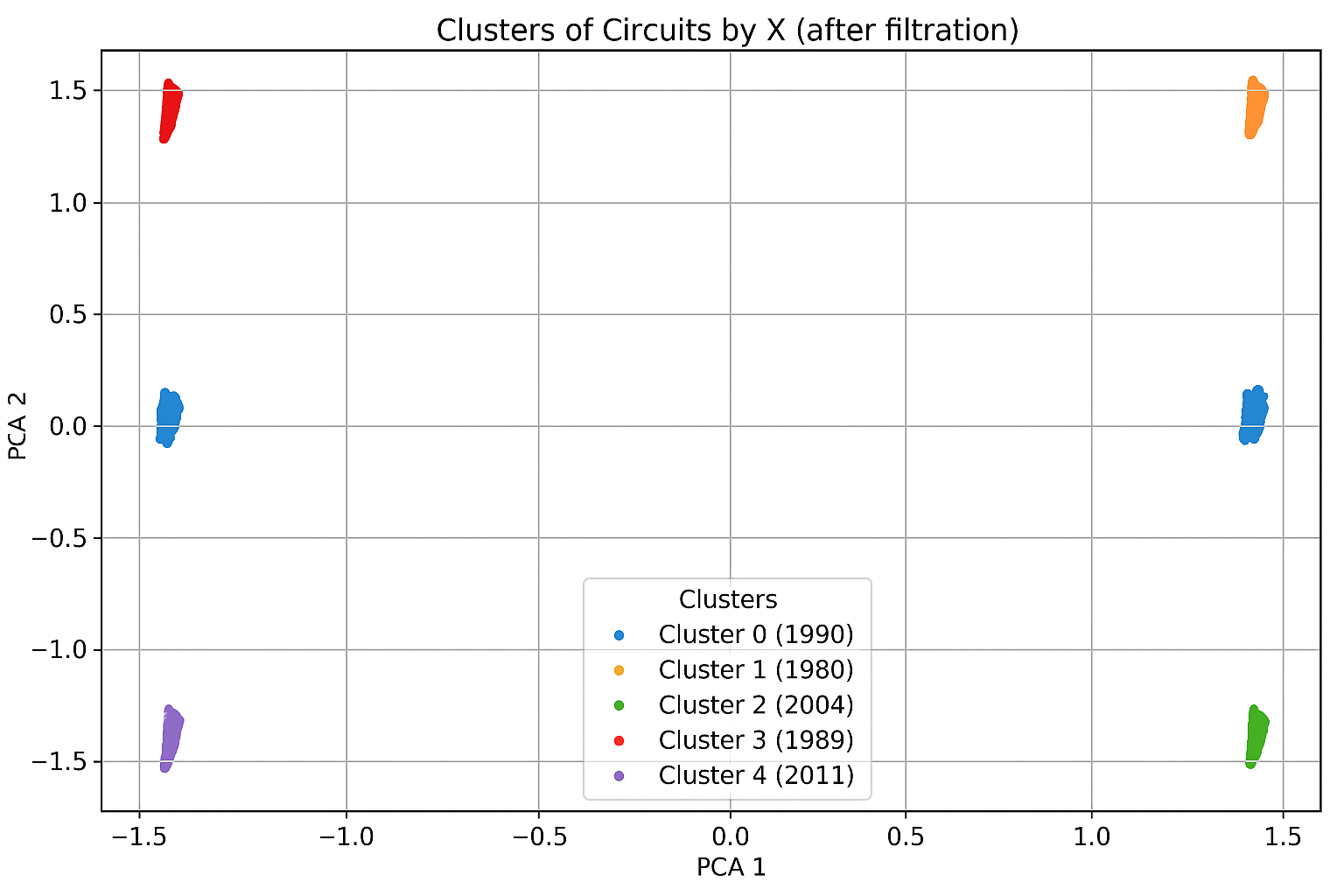}
        \caption{PCA visualization of the decompositions: each point is a circuit, and the clouds correspond to clusters of equivalent circuits.}
        \label{fig:cluster-positions}
    \end{minipage}
    \hfill
    \begin{minipage}{0.48\linewidth}
        \centering
        \includegraphics[trim=10 0 0 28,clip=true,width=1.05\linewidth]{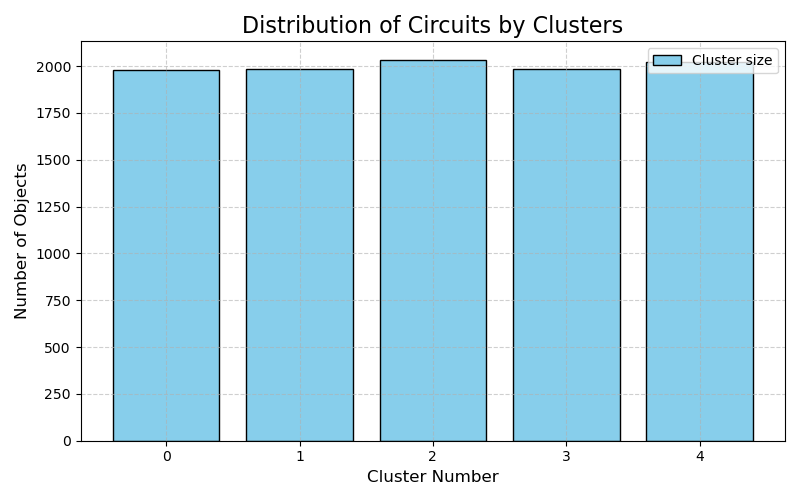}
        \caption{Number of circuits in each cluster.}
        \label{fig:cluster-counts}
    \end{minipage}
    \centering
    \includegraphics[width=.85\linewidth]{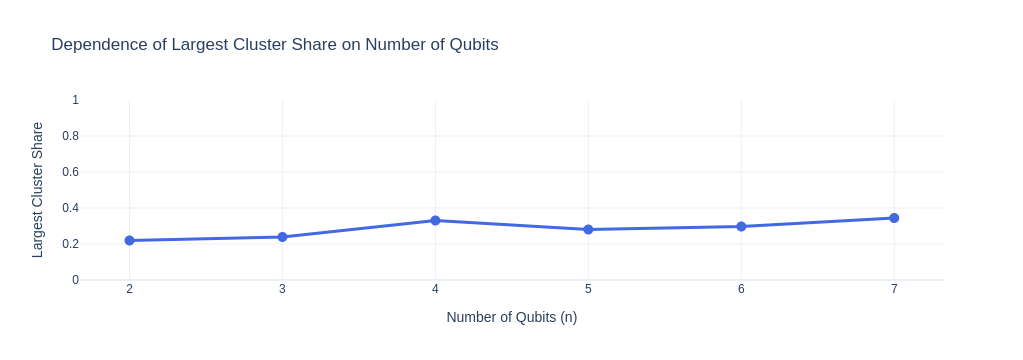}
    \caption{Dependence of Largest Cluster Share on Number of Qubits}
    \label{fig:cluster-max}
\end{figure}

We generate $10,000$ decompositions of two-qubit schemes with a diagonal operator.
In Fig.~\ref{fig:cluster-positions} and Fig.~\ref{fig:cluster-counts} we show the distribution of the number of objects per cluster.

From the resulting clusters, we can now take any one and search for a mapping for it. As can be seen in Fig. \ref{fig:cluster-max}, the share of useful data, generally speaking, does not depend on the number of qubits, which can be explained by the fact that of all possible schemes, the numerical decomposition methods fall into a limited number of patterns. However, the presence of several equivalent decomposition methods reduces the efficiency of data generation.

\paragraph{Training\label{sec:tp}}

We are working in a Euclidean $n$-dimensional space, that is, $\mathcal{X}=\RR^n$ is the parameter space of quantum circuits and $\mathcal{Y}=\RR^n$ is the parameter space of diagonal unitary operators. The main part of our mapping $\mathcal{A} =L_{lin}$ is linear and one-to-one, which means it does not contain noise, because it is generated by a theoretical, not an experimental dependence. Let $\gamma_k$  be the step of the gradient descent used to find $L_{lin}$ on our data, filtered and obtained in the \ref{sec:datagen1} section.
In our work, we use the following dependence of the gradient descent step on the step number $k$:

\begin{equation}     \{\gamma_k\} = \{\dots\underbrace{1/m^\alpha,\ 1/m^\alpha,\ \dots,\ 1/m^\alpha}_{\lceil m^\beta\rceil \text{ times}}, \underbrace{1/(m+1)^\alpha,\ 1/(m+1)^\alpha,\ \dots,\ 1/(m+1)^\alpha}_{\lceil(m+1)^\beta \rceil \text{ times}}\dots\}
\label{eq:sgd_step}
\end{equation}

where $\lceil x \rceil$ means the integer part of the number $x$ rounded up, which allows us to assume that the sum of the subsequences of the same values:

\begin{equation}
   \frac{1}{m^{\alpha-\beta}} \leqslant \sum_{i=1}^{\lceil m^\beta\rceil} \frac{1}{m^\alpha} \leqslant \frac{2}{m^{\alpha-\beta}}
    \label{eq:sgs_un_step}
\end{equation}

The \cite{sgd_prove} study indicates the necessary and sufficient conditions imposed on the problem to ensure convergence of SGD on linearly separable data.

Its authors prove the theorem that if we have an exactly specified (without variance) linear map (which, generally speaking, may turn out to be degenerate), then the SGD algorithm for linear model converges with probability equal to $1$, except for the set of measure $0$, or in other words the loss function is identically converted to $0$ if the following requirements are met:

\begin{enumerate}
    \item The input data does not contain noise, that is, the observations \(\mathbf{y}\) correspond exactly to the model \( \mathbf{y} = \mathcal{A} \mathbf{x} \).
  
    \item \( \mathcal{X} \)  is a Banach space that is strictly convex and smooth.

    \item The space allows for a dual mapping \( J : \mathcal{X} \to \mathcal{X}^* \), which is continuous and strictly monotonous.

    \item \( \mathcal{A} \) is linear and continuous operator \( \mathcal{A} : \mathcal{X} \to \mathcal{Y} \). 

    \item The operator \(\mathcal{A}\) has a closed image, which guarantees the existence of a solution with a minimum norm.

    \item The sequence \(\{ \gamma_k \} \subset \RR_+ \) is positive and satisfies the following conditions:
        \begin{equation}     \sum_{k=1}^{\infty} \gamma_k = \infty \quad \text{(divergence of the sum of steps)}, \end{equation}
        \begin{equation}     \sum_{k=1}^{\infty} \gamma_k^2 < \infty \quad \text{(convergence of the sum of square steps)}. \end{equation}

    \item The initial approximation of the linear mapping is chosen arbitrarily.

    \item At each iteration, a data element for SGD is randomly selected, which ensures the stochasticity of the method.

\end{enumerate}

\begin{figure}[h]
            \centering
            \includegraphics[width=0.75\linewidth]{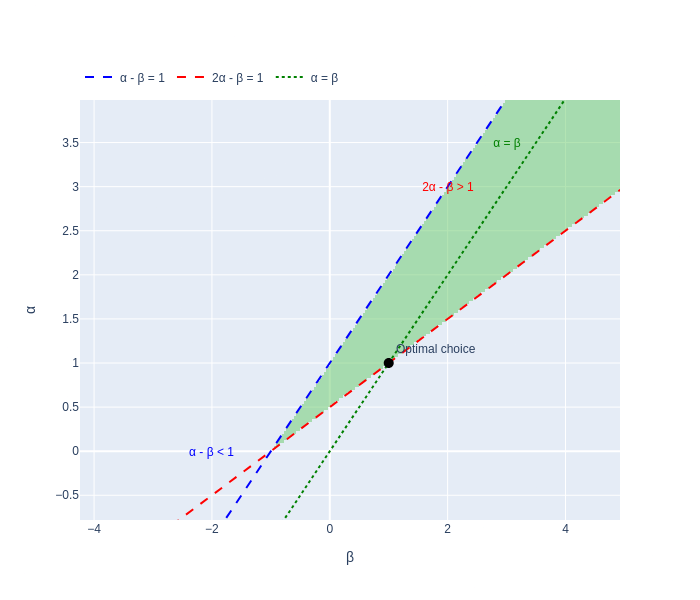}
            \caption{Plots of acceptable parameters $\alpha$ and $\beta$ and their optimal choice}
            \label{fig:optimal_lr}
        \end{figure}

In our situation, all the requirements except for $6$ are met, and the evidence is included in the Appendix  \ref{sec:Apendix}. 

To meet point $6$ restrictions should be imposed on $\alpha$ and $\beta$ (see Section \ref{sec:sgd_proove}):
\begin{equation}
\left\{
\begin{array}{rcl}
\alpha-\beta & \leqslant & 1 \\
2\alpha-\beta & > & 1 \\
\end{array}
\right. \Rightarrow
\left\{
\begin{array}{rcl}
\alpha & \leqslant & 1 + \beta \\
\alpha & > & \frac{1 + \beta}{2} \\
\end{array}
\right.
\end{equation}

Let's discuss the optimal choice of these parameters, see Fig. \ref{fig:optimal_lr}. Let the calculation time of one step of gradient descent be the same and independent of the step of gradient descent. This assumption is quite justified, because the SGD step is just a number in the formula. In this case, the larger the $\alpha$, the smaller our steps and the more steps to convergence, which means slower learning. On the other hand, the larger the $\beta$, the more steps of the same length, and therefore, for an equal number of iterations, the step of gradient descent decreases more slowly, which accelerates convergence. Thus, any choice along the straight line $\alpha = \beta$ will be optimal enough. In this case, the optimal choice, taking into account the constraints, will be $(1,1)$, graphically shown in Fig. \ref{fig:optimal_lr}.

Thus, if our model converges as a linear model, then a linear mapping exists, and if the model does not converge or converges as a semi-linear model, this means that a linear mapping does not exist - the only question is the speed of its convergence.

This sequence of gradient-descent steps is given in Algorithm~\ref{lst:HarmonicScheduler}.

\begin{algorithm}
\caption{Harmonic learning-rate schedule}\label{lst:HarmonicScheduler}
\begin{algorithmic}[1]
\Require initial rate $\eta_0$, block length $s$
\State $m\gets 1$, \quad $c\gets 0$ \Comment{$m$: current level; $c$: steps taken at this level}
\Function{NextRate}{ }
    \State $c\gets c+1$
    \If{$c\ge m\,s$}
        \State $m\gets m+1$, \quad $c\gets 0$
    \EndIf
    \State \Return $\eta_0/m$
\EndFunction
\end{algorithmic}
\end{algorithm}

The training of the model is summarized in Algorithm~\ref{lst:TrainingWorkflow_D}.

\begin{algorithm}
\caption{Training the model}\label{lst:TrainingWorkflow_D}
\begin{algorithmic}[1]
\Require dataset $\{(\vec\theta_i,\vec\varphi_i)\}$, model $L$ (Alg.~\ref{lst:linear_model}), schedule (Alg.~\ref{lst:HarmonicScheduler})
\For{$epoch=1,\dots,N_{\text{epochs}}$}
    \State $\hat{\vec\varphi}_i\gets L(\vec\theta_i)$ for all $i$
    \State $loss\gets \mathrm{RMSE}(\hat{\vec\varphi},\vec\varphi)$
    \State update the weights of $L$ by one stochastic gradient step on $loss$
    \State set the learning rate from the harmonic schedule
    \If{$loss<10^{-6}$}
        \State \textbf{break}
    \EndIf
\EndFor
\State \Return the weights of $L$: $W$ for linear part and $\Theta_g$ for non-linear
\end{algorithmic}
\end{algorithm}

\begin{figure}
    \centering
    \includegraphics[width=0.99\linewidth]{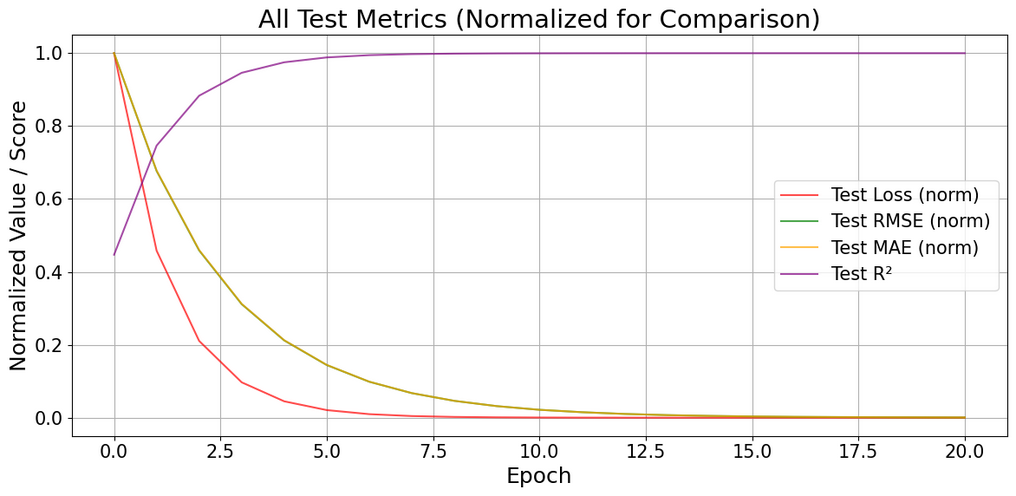}
    \caption{Normalized quality metrics and loss on the test set for the three-qubit Stage~1 model.}
    \label{fig:3_qubit_test_metric_D}
\end{figure}
In fact, the choice of hyperparameters depends on the number of qubits in the circuit that we are analyzing, but the algorithm itself remains unchanged.

\begin{figure}
    \centering
    \includegraphics[width=\linewidth]{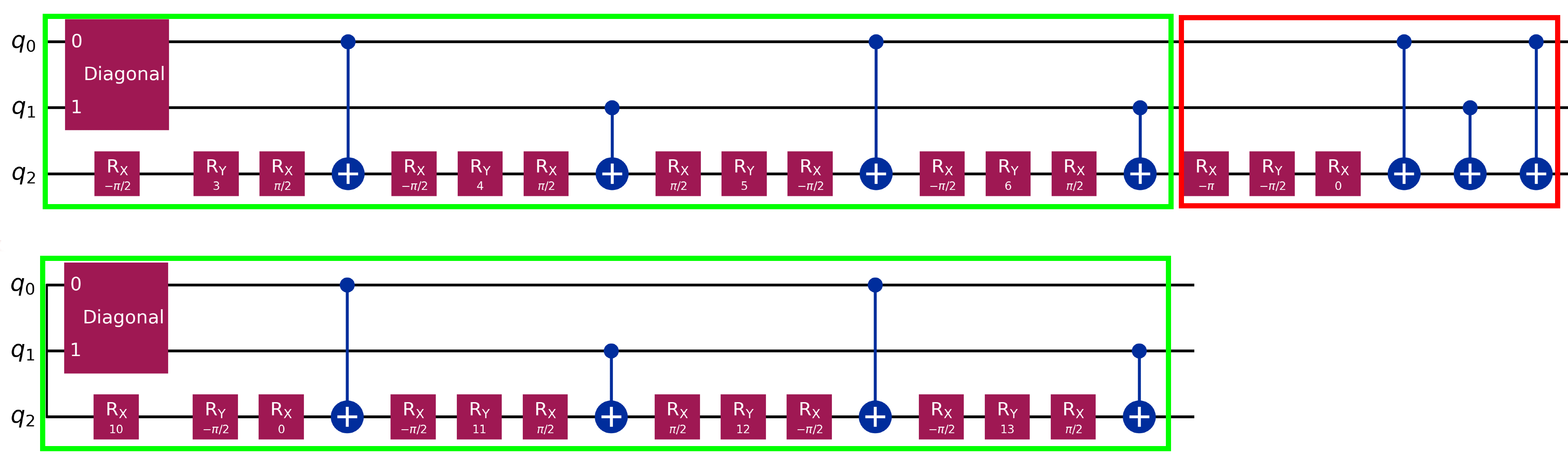}
    \caption{A three-qubit scheme generated by the decomposition routine}
    \label{fig:3q_qis}
    \centering
    \includegraphics[width=0.5\linewidth]{qc_optim_RZ_2_.png}
    \caption{A three-qubit scheme simplified by analyzing the resulting weight matrix of the trained model}
    \label{fig:3q_tor}
\end{figure}

As a result of applying this algorithm, we obtain a linear mapping matrix as the model weights, since the weights of the nonlinear part are vanishing. With good accuracy its entries take the values $\pm1$, together with some $\pm\frac12$ entries that come from the non-minimal circuits produced by the decomposition routine. These $\pm1$ weight matrices are the central object of this section: in Stage~$2$, where we work with a minimal ansatz, the half-integer entries disappear and the weights become the pure $\pm1$ matrices $r_n$ studied in the Math Hypothesis paragraph (Section~\ref{sec:tpf}). The graphs of metrics and the loss function on the validation sample are shown in Fig.~\ref{fig:3_qubit_test_metric_D}. The matrices are rectangular rather than square, again because the schemes found by the decomposition routine were generally not minimal. However, comparing the structure of the quantum circuit with the structure of the resulting matrix: highlighting dependent square submatrices, as well as constant structures (adding only a linear shift) we can significantly simplify the scheme, for example, as the three-qubit scheme in Fig. \ref{fig:3q_qis}. It can be simplified to Fig. \ref{fig:3q_tor}, using the symmetries of matrices and quantum circuits that we discussed earlier in \ref{sec:comp_exp}. In this case, the first part of the rectangular matrix and the last differ by a factor of $-1$ and, otherwise, coincide, and there are repeating sequences of operators in the scheme. Since all the quantum operators highlighted in rectangles are diagonal, and the one highlighted in the red rectangle gives only a parameter shift, we can leave only part of the operator. It remains to prove that it represents \textbf{all} three-qubit quantum diagonal operators. In general, for an arbitrary number of qubits, this proof is beyond the scope of this article and is discussed in detail in \cite{fedin2025mathematicalaspectsdecompositiondiagonal}.

\paragraph{Weights and QC comparison}

Looking at the weights of the model built using the non-optimized scheme generated by the decomposition routine, we see its block structure.

\begin{table*}[h!!!!!!!]
\centering
\small
\setlength{\arraycolsep}{2pt}
$W_{raw} = \left[\begin{array}{cccccccccccccc}
1 & 0 & 0 & -\frac{1}{2} & -\frac{1}{2} & -\frac{1}{2} & -\frac{1}{2} & 1 & 0 & 0 & -\frac{1}{2} & -\frac{1}{2} & -\frac{1}{2} & -\frac{1}{2} \\
0 & 1 & 0 & -\frac{1}{2} & \frac{1}{2} & \frac{1}{2} & -\frac{1}{2} & 0 & 1 & 0 & -\frac{1}{2} & \frac{1}{2} & \frac{1}{2} & -\frac{1}{2} \\
0 & 0 & 1 & -\frac{1}{2} & -\frac{1}{2} & \frac{1}{2} & \frac{1}{2} & 0 & 0 & 1 & -\frac{1}{2} & -\frac{1}{2} & \frac{1}{2} & \frac{1}{2} \\
-1 & -1 & -1 & -\frac{1}{2} & \frac{1}{2} & -\frac{1}{2} & \frac{1}{2} & -1 & -1 & -1 & -\frac{1}{2} & \frac{1}{2} & -\frac{1}{2} & \frac{1}{2} \\
1 & 0 & 0 & \frac{1}{2} & \frac{1}{2} & \frac{1}{2} & \frac{1}{2} & 1 & 0 & 0 & \frac{1}{2} & \frac{1}{2} & \frac{1}{2} & \frac{1}{2} \\
0 & 1 & 0 & \frac{1}{2} & -\frac{1}{2} & -\frac{1}{2} & \frac{1}{2} & 0 & 1 & 0 & \frac{1}{2} & -\frac{1}{2} & -\frac{1}{2} & \frac{1}{2} \\
0 & 0 & 1 & \frac{1}{2} & \frac{1}{2} & -\frac{1}{2} & -\frac{1}{2} & 0 & 0 & 1 & \frac{1}{2} & \frac{1}{2} & -\frac{1}{2} & -\frac{1}{2} \\
-1 & -1 & -1 & \frac{1}{2} & -\frac{1}{2} & \frac{1}{2} & -\frac{1}{2} & -1 & -1 & -1 & \frac{1}{2} & -\frac{1}{2} & \frac{1}{2} & -\frac{1}{2}
\end{array}\right];\qquad$
$b_{raw} = \left[\begin{array}{cccccccc}
-0.838 & -0.838 & -0.209 & -0.209 & 0.838 & 0.838 & 0.209 & 0.209
\end{array}\right]$
\end{table*}

\begin{table*}[h!!!!!!!]
\centering
\small
\setlength{\arraycolsep}{2pt}
$W = \left[\begin{array}{ccccccc}
1 & 0 & 0 & -\frac{1}{2} & -\frac{1}{2} & -\frac{1}{2} & -\frac{1}{2} \\
0 & 1 & 0 & -\frac{1}{2} & \frac{1}{2} & \frac{1}{2} & -\frac{1}{2} \\
0 & 0 & 1 & -\frac{1}{2} & -\frac{1}{2} & \frac{1}{2} & \frac{1}{2} \\
-1 & -1 & -1 & -\frac{1}{2} & \frac{1}{2} & -\frac{1}{2} & \frac{1}{2} \\
1 & 0 & 0 & \frac{1}{2} & \frac{1}{2} & \frac{1}{2} & \frac{1}{2} \\
0 & 1 & 0 & \frac{1}{2} & -\frac{1}{2} & -\frac{1}{2} & \frac{1}{2} \\
0 & 0 & 1 & \frac{1}{2} & \frac{1}{2} & -\frac{1}{2} & -\frac{1}{2} \\
-1 & -1 & -1 & \frac{1}{2} & -\frac{1}{2} & \frac{1}{2} & -\frac{1}{2} \\
\end{array}\right];\qquad$
$b = \left[\begin{array}{cccccccc}
0 &0 &0 &0 &0 &0 &0&0 \\
\end{array}\right]$
\end{table*}

We can write that $W_{raw}$ has the form of a block matrix:

\begin{equation}
    W_{raw} = \begin{bmatrix}
        W&W\\
    \end{bmatrix}
\end{equation}
The vector $b_{raw}$ follows from the structure in the middle of the quantum circuit (Fig.~\ref{fig:3q_qis}): it gives a constant phase shift of the diagonal elements and does not affect the solution. Note that $W$ still contains $\pm\frac12$ entries because the routine's circuit is not minimal; it is the Stage $1$ counterpart of the pure $\pm1$ matrices $r_n$ that we obtain from the minimal ansatz in Stage~$2$.

\paragraph{Further options for generating circuits}

Based on the schemes obtained in the previous part for two and three qubits, we can assume two possible sequences corresponding to the numbers of the control qubits $A_{\cdot,n}$, which determine the order of application of the $CNOT$ tail gates in diagonal quantum circuits. Both methods are presented below.

\subparagraph{Strange Fractal}
This approach considers the possibility of representing the $CNOT$ tail as two identical operators with different parameters. 
This suggests that the sequence $A_{\cdot,n}$ can be constructed iteratively based on the previous steps. 

\begin{figure}[htbp]
    \centering
    \begin{minipage}{0.55\linewidth}
        \centering
        \includegraphics[trim = 10 0 10 0, clip = true, width=\linewidth]{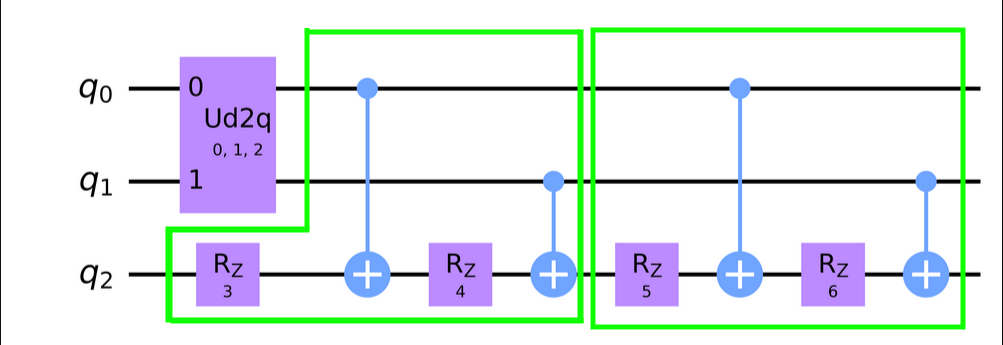}
    \end{minipage}
    \hfill
    \begin{minipage}{0.43\linewidth}
        \centering
        \includegraphics[width=\linewidth]{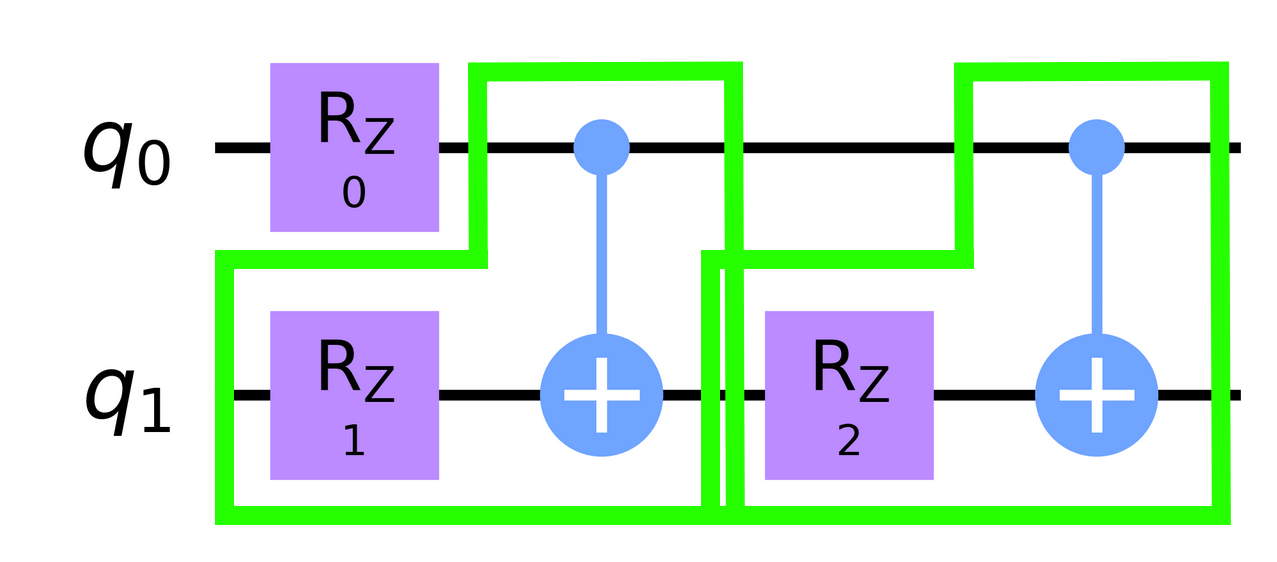}
    \end{minipage}
    \caption{A variant of splitting three-qubit and two-qubit circuits.}
    \label{fig:2qv1}
\end{figure}

To do this, let's recall Fig. \ref{fig:simple_2qubit} and Fig. \ref{fig:3q_tor}. Let's break down the scheme as follows: after the diagonal operator obtained in the previous step, we add two structurally identical but parametrically different operators, see Fig. \ref{fig:2qv1}.

To do this, we introduce the concatenation operation $\circ$ for sequences $a_\cdot$ and $b_\cdot$:

\begin{equation}
     \{a_1, a_2, \dots a_N\} \circ \{b_1, b_2, \dots b_M\} = \{a_1, a_2, \dots a_N, b_1, b_2, \dots b_M\}
    \label{eq:concat_definition}
\end{equation}

We also use the notation of ordinary arithmetic expressions over sequences:

\begin{equation}
    \begin{aligned}
        a_{\cdot} + m &= \{a_1, a_2, \dots, a_N\} + m = \{a_1 + m, a_2 + m, \dots, a_N + m\}; \\
        a_{\cdot} \cdot m &= \{a_1, a_2, \dots, a_N\} \cdot m = \{a_1 \cdot m, a_2 \cdot m, \dots, a_N \cdot m\}
    \end{aligned}
    \label{eq:cons_def_math}
\end{equation}

Then the sequence is defined as:
\begin{equation}
A_{\cdot,n} = a_{\cdot,n} \circ a_{\cdot,n},
\end{equation}
where $a_{\cdot,n}$ is half of the desired sequence, sufficient to generate $A_{\cdot,n}$.

In the first steps, which correspond to the two- and three-qubit schemes obtained earlier in \ref{sec:tp}
\begin{equation}
a_{\cdot,2} = \{1\}, \quad a_{\cdot,3} = \{1,2\}.
\end{equation}
Note that:
\begin{equation}
a_{\cdot,3} = a_{\cdot,2} \circ (3 - a_{\cdot,2}).
\end{equation}
This allows us to generalize the construction to the next step:\begin{equation}
a_{\cdot,n} = a_{\cdot,n-1} \circ (n - a_{\cdot,n-1}).
\end{equation}
In this case, we get:
\begin{equation}
\begin{aligned}
a_{\cdot,3} &= \{1,2\},\\
a_{\cdot,4} &= \{1,2,3,2\},\\
a_{\cdot,5} &= \{1,2,3,2,4,3,2,3\}.
\end{aligned}
\end{equation}
Thus, the structure of each new sequence is constructed as a symmetrical deformation of the previous one. Checking the degeneracy of the corresponding matrices in specific cases (for example, through the determinant) shows that these schemes are suitable for describing diagonal unitary operators for each special case.

\subparagraph{Binary Tree}

\begin{figure}[htbp]
    \centering
    \begin{minipage}{0.55\linewidth}
        \centering
        \includegraphics[trim = 10 0 10 0, clip = true, width=\linewidth]{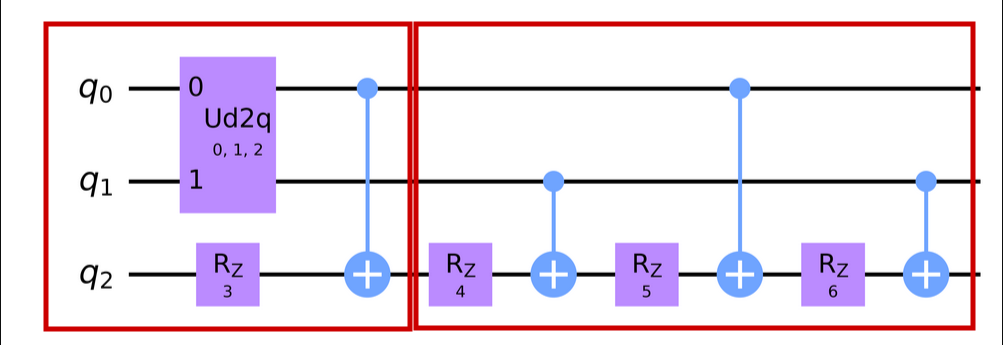}
    \end{minipage}
    \hfill
    \begin{minipage}{0.43\linewidth}
        \centering
        \includegraphics[width=\linewidth]{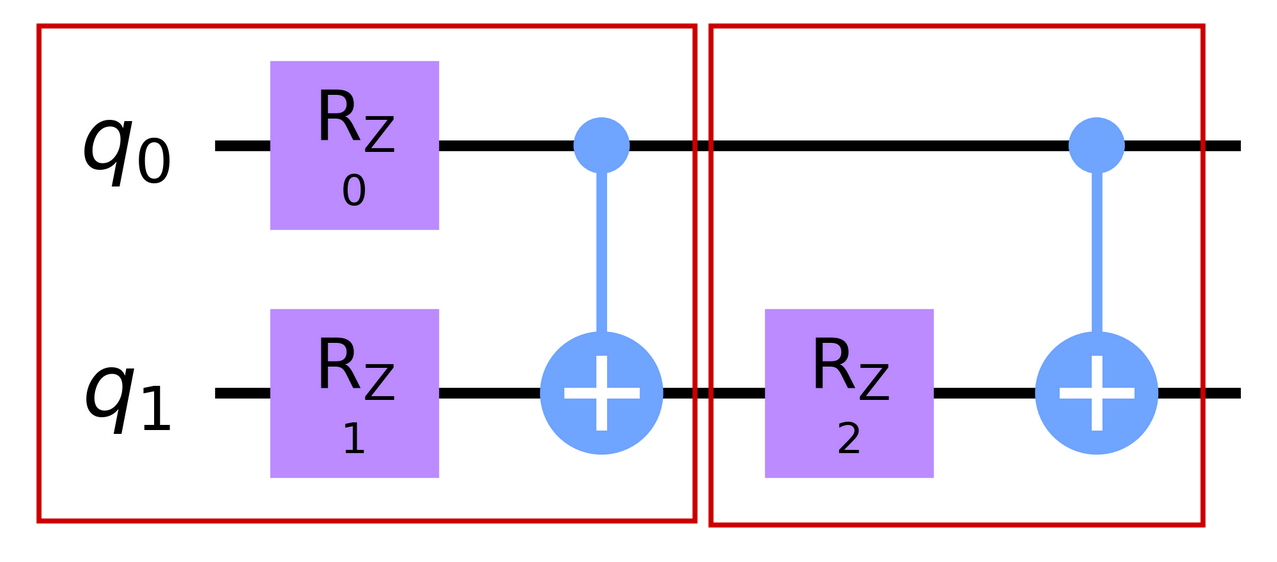}
    \end{minipage}
    \caption{Another variant of splitting three-qubit and two-qubit circuits.}
    \label{fig:3qv2}

\end{figure}

The second approach is based on the analysis of the scheme structure in the form of a tree (see Fig. \ref{fig:3qv2}). This approach leads to a different kind of recursive construction:
\begin{equation}
A_{\cdot,n} = \{1\} \circ a_{\cdot,n},
\label{fractal_sequence}
\end{equation}
where the auxiliary sequence $a_{\cdot,n}$ is defined as:
\begin{equation}
a_{\cdot,n} = (a_{\cdot,n-1} + 1) \circ \{1\} \circ (a_{\cdot,n-1} + 1).
\label{fractal_sequence_aux}
\end{equation}

This approach leads to a self-similar (fractal) structure of sequences and is simpler in terms of provability, since it reflects the symmetric nesting of operators at each level of the scheme. This fractal is called a binary tree.

 \begin{definition}(Perfect Binary Tree (PBT)\label{def:PBT})
The set $\mathcal{P}_h$ of perfect binary trees of height $h \in \mathbb{N}$ is defined recursively. A tree is a pair $T=(V,E)$ of vertices and edges.
For $h=1$, $\mathcal{P}_1$ contains the single tree $T_1 = (\{v_1\}, \emptyset)$.
For $h>1$, a tree $T=(V,E) \in \mathcal{P}_h$ is formed from a new root vertex $v$ and two trees $T_L=(V_L, E_L)$ and $T_R=(V_R, E_R)$ from $\mathcal{P}_{h-1}$ with $V_L \cap V_R = \emptyset$, such that
\begin{align}
    V &= V_L \cup \{v\} \cup V_R, \\
    E &= E_L \cup \{ (v, v_L), (v, v_R) \} \cup E_R ,
\end{align}
where $v_L$ and $v_R$ are the roots of $T_L$ and $T_R$ respectively.
\end{definition}

The graphical representation of this is presented on Fig. \ref{fig:PBT_steps} and more details about binary trees can be found in \cite{harder_ece250_pbt} and \cite{Knuth1997TAOCP1}. An in-depth study of this case and a proof of the generality of such a decomposition, as well as its possible symmetries, are discussed in \cite{fedin2025mathematicalaspectsdecompositiondiagonal}.

\begin{figure}
\centering
\scalebox{0.75}{%
\begin{tabular}{cccc}
\begin{tikzpicture}[arrowstyle/.style={-{Stealth[length=2mm,width=2mm]}, thick, draw=black!70}]
  \node[treenode] (root) {$1$};
  \coordinate (tRoot) at ($(root.south)+(0,-10mm)$);
  \draw[arrowstyle] (root.south) -- (tRoot);
  \node[anchor=north] at (tRoot) {$\{1\}$};
\end{tikzpicture}

&
\begin{tikzpicture}[level distance=10mm,
  level 1/.style={sibling distance=10mm},
  arrowstyle/.style={-{Stealth[length=2mm,width=2mm]}, thick, draw=black!70}
  ]
  \node[treenode] (root) {$1$}
    child {node[treenode] (L) {$2$}}
    child {node[treenode] (R) {$2$}};
  \coordinate (tL)    at ($(L.south)+(0,-7mm)$);
  \coordinate (tRoot) at ($(root.south)+(0,-17mm)$);
  \coordinate (tR)    at ($(R.south)+(0,-7mm)$);
  \draw[arrowstyle] (L.south)    -- (tL);
  \draw[arrowstyle] (root.south) -- (tRoot);
  \draw[arrowstyle] (R.south)    -- (tR);
  \node[anchor=north] at (tL)    {$\{2,$};
  \node[anchor=north] at (tRoot) {$1$};
  \node[anchor=north] at (tR)    {$, 2\}$};
\end{tikzpicture}

&
\begin{tikzpicture}[level distance=10mm,
  level 1/.style={sibling distance=20mm},
  level 2/.style={sibling distance=10mm},
  arrowstyle/.style={-{Stealth[length=2mm,width=2mm]}, thick, draw=black!70}
  ]

  \node[treenode] (root) {$1$}
    child {node[treenode] (L) {$2$}
      child {node[treenode] (LL) {$3$}}
      child {node[treenode] (LR) {$3$}}
    }
    child {node[treenode] (R) {$2$}
      child {node[treenode] (RL) {$3$}}
      child {node[treenode] (RR) {$3$}}
    };
  \coordinate (tLL) at ($(LL.south)+(0,-7mm)$);
  \coordinate (tLR) at ($(LR.south)+(0,-7mm)$);
  \coordinate (tL)  at ($(L.south)+(0,-17mm)$);
  \coordinate (tR)  at ($(R.south)+(0,-17mm)$);
  \coordinate (tRL) at ($(RL.south)+(0,-7mm)$);
  \coordinate (tRR) at ($(RR.south)+(0,-7mm)$);
  \coordinate (tRoot) at ($(root.south)+(0,-27mm)$);
  \draw[arrowstyle] (LL.south) -- (tLL);
  \draw[arrowstyle] (LR.south) -- (tLR);
  \draw[arrowstyle] (L.south)  -- (tL);
  \draw[arrowstyle] (R.south)  -- (tR);
  \draw[arrowstyle] (RL.south) -- (tRL);
  \draw[arrowstyle] (RR.south) -- (tRR);
  \draw[arrowstyle] (root.south) -- (tRoot);
  \node[anchor=north] at (tLL) {$\{3, $};
  \node[anchor=north] at (tLR) {$3, $};
  \node[anchor=north] at (tL)  {$2, $};
  \node[anchor=north] at (tR)  {$2, $};
  \node[anchor=north] at (tRL) {$3,$};
  \node[anchor=north] at (tRR) {$3\}$};
  \node[anchor=north] at (tRoot) {$ 1,$};
\end{tikzpicture}

&
\begin{tikzpicture}[level distance=10mm,
  level 1/.style={sibling distance=40mm},
  level 2/.style={sibling distance=20mm},
  level 3/.style={sibling distance=10mm},
  arrowstyle/.style={-{Stealth[length=2mm,width=2mm]}, thick, draw=black!70}
  ]
  \node[treenode] (root) {$1$}
    child {node[treenode] (L) {$2$}
      child {node[treenode] (L1) {$3$}
        child {node[treenode] (L11) {$4$}}
        child {node[treenode] (L12) {$4$}}
      }
      child {node[treenode] (L2) {$3$}
        child {node[treenode] (L21) {$4$}}
        child {node[treenode] (L22) {$4$}}
      }
    }
    child {node[treenode] (R) {$2$}
      child {node[treenode] (R1) {$3$}
        child {node[treenode] (R11) {$4$}}
        child {node[treenode] (R12) {$4$}}
      }
      child {node[treenode] (R2) {$3$}
        child {node[treenode] (R21) {$4$}}
        child {node[treenode] (R22) {$4$}}
      }
    };
  \coordinate (tL11) at ($(L11.south)+(0,-7mm)$);
  \coordinate (tL12) at ($(L12.south)+(0,-7mm)$);
  \coordinate (tL21) at ($(L21.south)+(0,-7mm)$);
  \coordinate (tL22) at ($(L22.south)+(0,-7mm)$);
  \coordinate (tR11) at ($(R11.south)+(0,-7mm)$);
  \coordinate (tR12) at ($(R12.south)+(0,-7mm)$);
  \coordinate (tR21) at ($(R21.south)+(0,-7mm)$);
  \coordinate (tR22) at ($(R22.south)+(0,-7mm)$);

  \coordinate (tL1) at ($(L1.south)+(0,-17mm)$);
  \coordinate (tL2) at ($(L2.south)+(0,-17mm)$);
  \coordinate (tR1) at ($(R1.south)+(0,-17mm)$);
  \coordinate (tR2) at ($(R2.south)+(0,-17mm)$);

  \coordinate (tL)  at ($(L.south)+(0,-27mm)$);
  \coordinate (tR)  at ($(R.south)+(0,-27mm)$);
  \coordinate (tRoot) at ($(root.south)+(0,-37mm)$);
  \foreach \a/\b in {L11/tL11,L12/tL12,L21/tL21,L22/tL22,
                     R11/tR11,R12/tR12,R21/tR21,R22/tR22,
                     L1/tL1,L2/tL2,R1/tR1,R2/tR2,
                     L/tL,R/tR,root/tRoot}
    \draw[arrowstyle] (\a.south) -- (\b);
  \node[anchor=north] at (tL11) {$\{4,$};
  \node[anchor=north] at (tL12) {$4,$};
  \node[anchor=north] at (tL21) {$4,$};
  \node[anchor=north] at (tL22) {$4,$};

  \node[anchor=north] at (tR11) {$4,$};
  \node[anchor=north] at (tR12) {$4,$};
  \node[anchor=north] at (tR21) {$4,$};
  \node[anchor=north] at (tR22) {$4\}$};

  \node[anchor=north] at (tL1) {$3,$};
  \node[anchor=north] at (tL2) {$3,$};
  \node[anchor=north] at (tR1) {$3,$};
  \node[anchor=north] at (tR2) {$3,$};

  \node[anchor=north] at (tL)  {$2,$};
  \node[anchor=north] at (tR)  {$2,$};

  \node[anchor=north] at (tRoot) {$ 1,$};

\end{tikzpicture}

\\[2em]

\multicolumn{4}{c}{
\begin{tikzpicture}[level distance=10mm,
  level 1/.style={sibling distance=80mm},
  level 2/.style={sibling distance=40mm},
  level 3/.style={sibling distance=20mm},
  level 4/.style={sibling distance=10mm},
  arrowstyle/.style={-{Stealth[length=2mm,width=2mm]}, thick, draw=black!70}
  ]

  \node[treenode] (root) {$1$}
    child {node[treenode] (L) {$2$}
      child {node[treenode] (L1) {$3$}
        child {node[treenode] (L11) {$4$}
          child {node[treenode] (L111) {$5$}}
          child {node[treenode] (L112) {$5$}}
        }
        child {node[treenode] (L12) {$4$}
          child {node[treenode] (L121) {$5$}}
          child {node[treenode] (L122) {$5$}}
        }
      }
      child {node[treenode] (L2) {$3$}
        child {node[treenode] (L21) {$4$}
          child {node[treenode] (L211) {$5$}}
          child {node[treenode] (L212) {$5$}}
        }
        child {node[treenode] (L22) {$4$}
          child {node[treenode] (L221) {$5$}}
          child {node[treenode] (L222) {$5$}}
        }
      }
    }
    child {node[treenode] (R) {$2$}
      child {node[treenode] (R1) {$3$}
        child {node[treenode] (R11) {$4$}
          child {node[treenode] (R111) {$5$}}
          child {node[treenode] (R112) {$5$}}
        }
        child {node[treenode] (R12) {$4$}
          child {node[treenode] (R121) {$5$}}
          child {node[treenode] (R122) {$5$}}
        }
      }
      child {node[treenode] (R2) {$3$}
        child {node[treenode] (R21) {$4$}
          child {node[treenode] (R211) {$5$}}
          child {node[treenode] (R212) {$5$}}
        }
        child {node[treenode] (R22) {$4$}
          child {node[treenode] (R221) {$5$}}
          child {node[treenode] (R222) {$5$}}
        }
      }
    };
  \foreach \n/\c in {L111/tL111,L112/tL112,L121/tL121,L122/tL122,
                     L211/tL211,L212/tL212,L221/tL221,L222/tL222,
                     R111/tR111,R112/tR112,R121/tR121,R122/tR122,
                     R211/tR211,R212/tR212,R221/tR221,R222/tR222}
    \coordinate (\c) at ($(\n.south)+(0,-7mm)$);

  \foreach \n/\c in {L11/tL11,L12/tL12,L21/tL21,L22/tL22,
                     R11/tR11,R12/tR12,R21/tR21,R22/tR22}
    \coordinate (\c) at ($(\n.south)+(0,-17mm)$);

  \foreach \n/\c in {L1/tL1,L2/tL2,R1/tR1,R2/tR2}
    \coordinate (\c) at ($(\n.south)+(0,-27mm)$);

  \foreach \n/\c in {L/tL,R/tR}
    \coordinate (\c) at ($(\n.south)+(0,-37mm)$);
  \coordinate (tRoot) at ($(root.south)+(0,-47mm)$);

  \foreach \a/\b in {L111/tL111,L112/tL112,L121/tL121,L122/tL122,
                     L211/tL211,L212/tL212,L221/tL221,L222/tL222,
                     R111/tR111,R112/tR112,R121/tR121,R122/tR122,
                     R211/tR211,R212/tR212,R221/tR221,R222/tR222,
                     L11/tL11,L12/tL12,L21/tL21,L22/tL22,
                     R11/tR11,R12/tR12,R21/tR21,R22/tR22,
                     L1/tL1,L2/tL2,R1/tR1,R2/tR2,
                     L/tL,R/tR,root/tRoot}
    \draw[arrowstyle] (\a.south) -- (\b);

  \node[anchor=north] at (tL111) {$\{5,$};
  \node[anchor=north] at (tL112) {$5,$};
  \node[anchor=north] at (tL121) {$5,$};
  \node[anchor=north] at (tL122) {$5,$};
  \node[anchor=north] at (tL211) {$5,$};
  \node[anchor=north] at (tL212) {$5,$};
  \node[anchor=north] at (tL221) {$5,$};
  \node[anchor=north] at (tL222) {$5,$};

  \node[anchor=north] at (tR111) {$5,$};
  \node[anchor=north] at (tR112) {$5,$};
  \node[anchor=north] at (tR121) {$5,$};
  \node[anchor=north] at (tR122) {$5,$};
  \node[anchor=north] at (tR211) {$5,$};
  \node[anchor=north] at (tR212) {$5,$};
  \node[anchor=north] at (tR221) {$5,$};
  \node[anchor=north] at (tR222) {$5\}$};

  \node[anchor=north] at (tL11) {$4,$};
  \node[anchor=north] at (tL12) {$4,$};
  \node[anchor=north] at (tL21) {$4,$};
  \node[anchor=north] at (tL22) {$4,$};
  \node[anchor=north] at (tR11) {$4,$};
  \node[anchor=north] at (tR12) {$4,$};
  \node[anchor=north] at (tR21) {$4,$};
  \node[anchor=north] at (tR22) {$4,$};

  \node[anchor=north] at (tL1) {$3,$};
  \node[anchor=north] at (tL2) {$3,$};
  \node[anchor=north] at (tR1) {$3,$};
  \node[anchor=north] at (tR2) {$3,$};

  \node[anchor=north] at (tL) {$2,$};
  \node[anchor=north] at (tR) {$2,$};

  \node[anchor=north] at (tRoot) {$1,$};

\end{tikzpicture}
}
\end{tabular}
}
\caption{Perfect binary trees of heights $1$ to $5$, with the corresponding control-qubit sequence shown beneath each tree.}
\label{fig:PBT_steps}
\end{figure}

Both schemes can be used depending on the symmetry requirements, computational complexity, and interpretability of the quantum circuit.

\subsubsection{Stage \texorpdfstring{$2$}{2}}

In this Stage of the workflow, starting from the simplified circuit found in Stage $1$, we generate Pretty Data from the ansatz, train the model with the same architecture as the one in a previous Stage, and analyze its weights (the MWA step) to obtain the mathematical hypothesis.

\begin{figure}
    \centering
    \includegraphics[width=\linewidth]{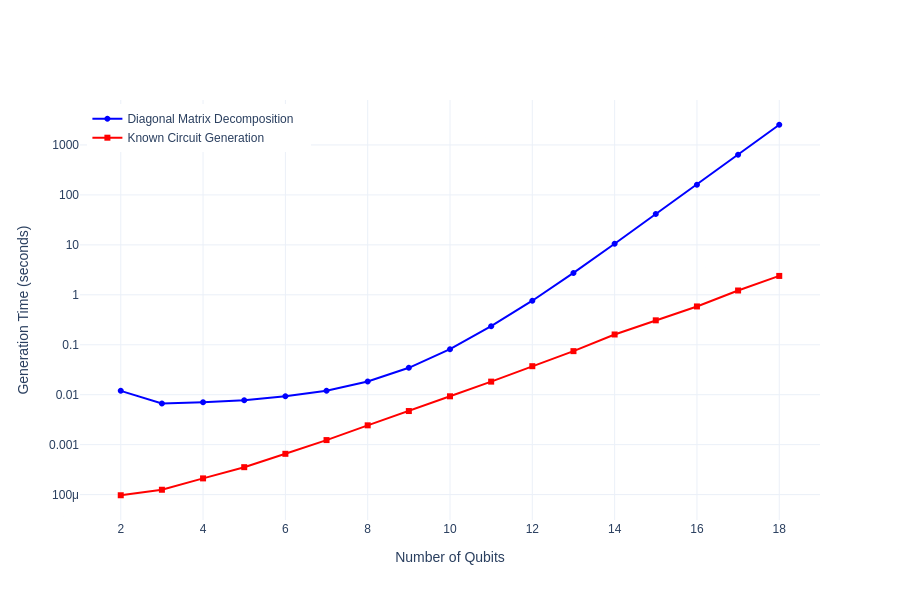}
    \caption{Comparison of the time spent on decomposition with the generation time of a known circuit, depending on the number of qubits in the circuit.}
    \label{fig:decomp_vs_gen}
\end{figure}

\paragraph{Dataset \label{sec:datagen2}}
Now we have an assumption based on the first three orders of decomposition of the $n-$qubit circuit. We will generate multi-qubit circuits based on this circuit. This is computationally advantageous, since generating a circuit of a given order and evaluating its unitary matrix takes far less time than the reverse process - decomposing that unitary matrix into a circuit. Fig. \ref{fig:decomp_vs_gen} clearly compares the generation time of a given circuit with the decomposition time of a diagonal operator obtained via the built-in decomposition routine.

The construction of the ansatz circuit, which follows the binary-tree sequence of the recursive formula~\eqref{fractal_sequence}, is given in Algorithm~\ref{lst:general_qc_D}.

\begin{algorithm}
\caption{Recursive ansatz for the diagonal $n$-qubit operator}\label{lst:general_qc_D}
\begin{algorithmic}[1]
\Function{BuildDiagonal}{$n$, parameters $\vec p$}
    \If{$n=1$}
        \State \Return a circuit with a single $R_Z(p_1)$ on the qubit
    \EndIf
    \If{$n=2$}
        \State \Return the two-qubit block of Fig.~\ref{fig:simple_2qubit} with parameters $p_1,p_2,p_3$
    \EndIf
    \State build \Call{BuildDiagonal}{$n-1$, first $2^{\,n-1}-1$ parameters} on qubits $1,\dots,n-1$
    \State $S\gets$ control-qubit sequence for $n$ qubits, given by the binary tree of Eq.~\eqref{fractal_sequence}
    \For{pass $=1,2$}
        \For{each control $c\in S$}
            \State apply $R_Z$ with the next parameter on qubit $n$
            \State apply a $CNOT$ with control $c$ and target $n$
        \EndFor
    \EndFor
    \State \Return the assembled circuit
\EndFunction
\end{algorithmic}
\end{algorithm}

\paragraph{Training}

By generating circuits for different numbers of qubits, we train the model discussed above. The process of learning a model in this case is no different from \ref{sec:tp}. 

It is worth noting here that we are generating data from a fairly small $n-$ dimensional hypercube with a center of $0$ and a side of $\epsilon$. The symmetry of the cube relative to the point $0$ is necessary, because if there is a large number of $+1$ in one line in our display, the parameter can be moved beyond $[-\pi,\pi]$, in this case the display will cease to be one-to-one and our algorithm will not be able to converge. 
For specific experiments, the side of the cube is $2\epsilon = 0.05$, that is, we use $U_{0.025}(0)$, see Section \ref{sec:gen_pretty_date}.

\begin{figure}[h!]
    \centering
    \begin{minipage}{0.495\textwidth}

                \centering
        \includegraphics[width=\linewidth]{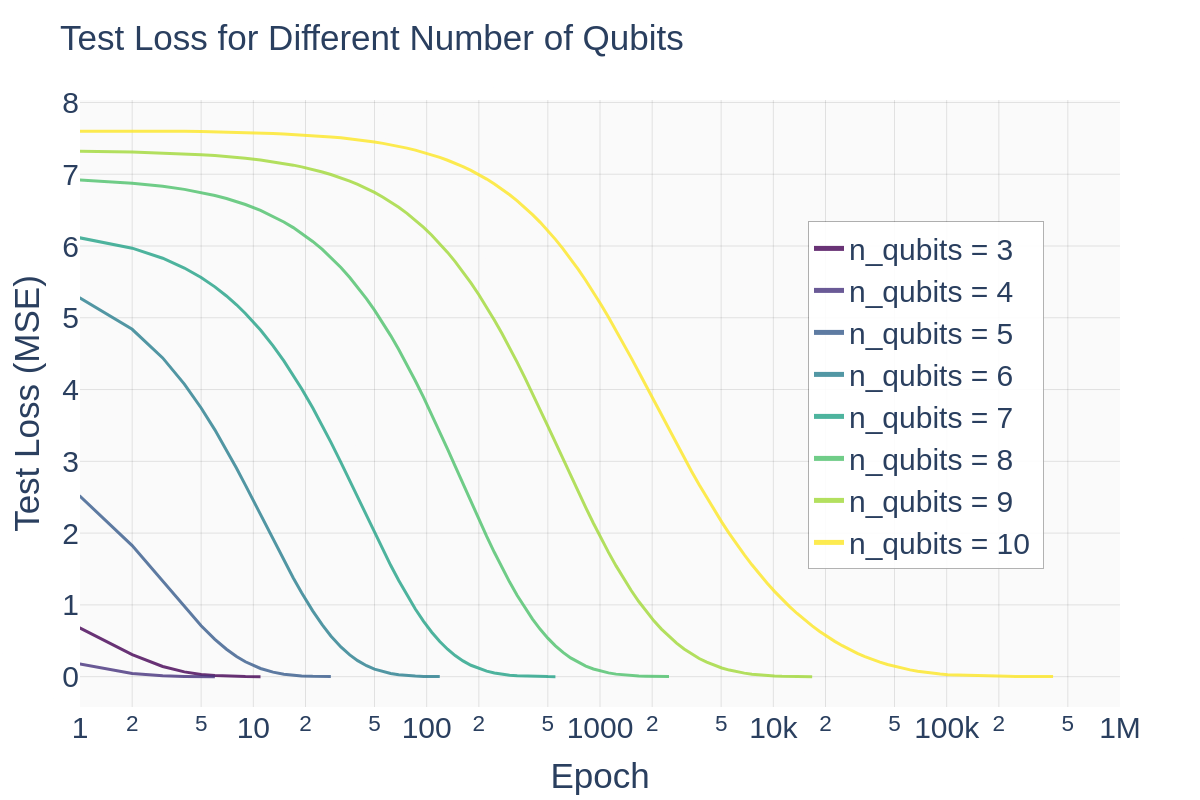}
         \caption{Test loss for different numbers of qubits.}
        \label{fig:loss_results}
    \end{minipage}
    \hfill
    \begin{minipage}{0.495\textwidth}
        \centering
        \includegraphics[width=\linewidth]{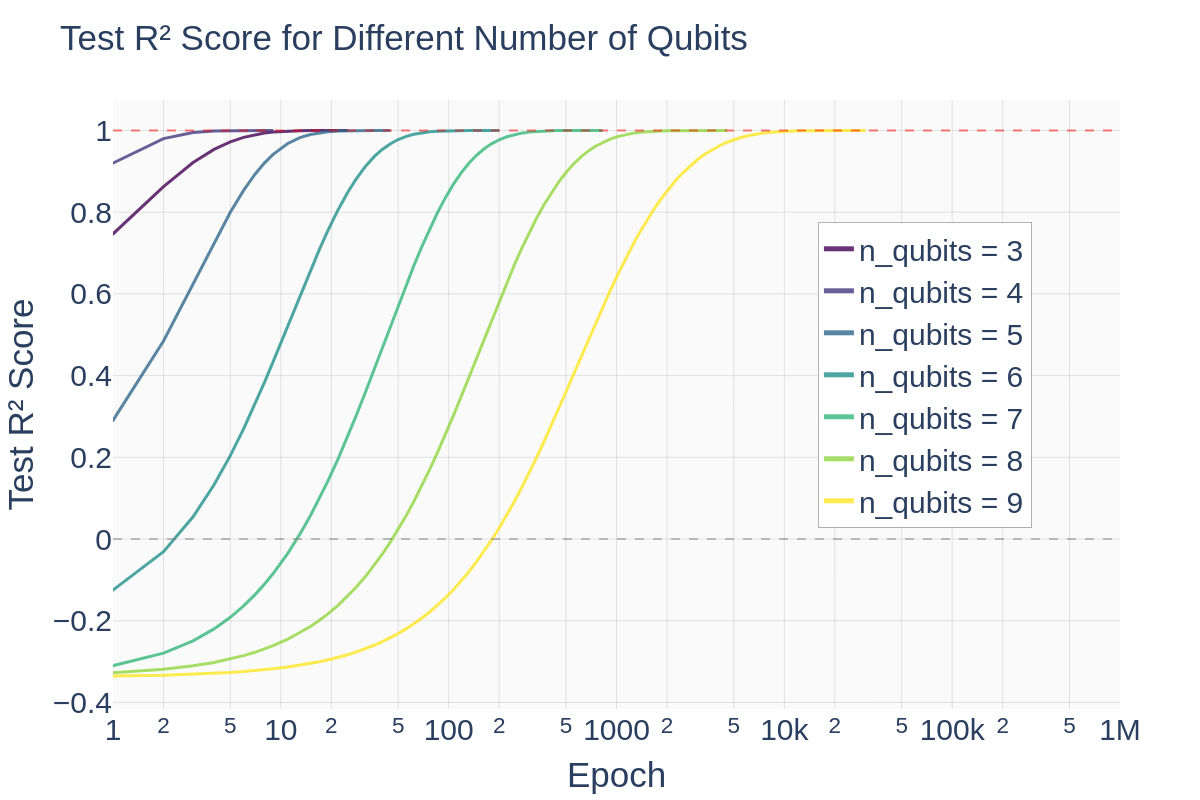}
        \caption{R² score for different numbers of qubits}
        \label{fig:r2_results}
    \end{minipage}
\end{figure}

\begin{figure}[h!]
    \centering
    \includegraphics[width=0.75\linewidth]{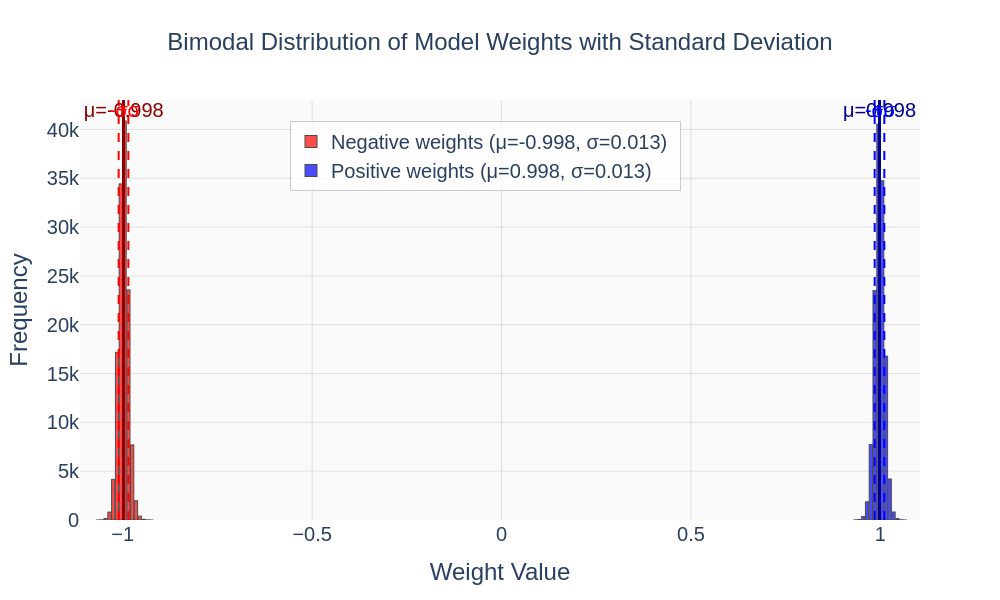}
    \caption{Distribution of the model weights for $n = 10$: the values cluster tightly around $+1$ and $-1$.}
    \label{fig:weigts_distribution}
\end{figure}

Although the model was semi-linear, its nonlinear part vanished completely during training under a small regularization penalty for all tested $n$, confirming that the linearity was not imposed a priori. When analyzing the weights of the model for $n = 10$, we obtain the distribution of Fig.~\ref{fig:weigts_distribution}. It agrees perfectly with our assumption that the entries of the weight matrix take only two values, which are very well separated. These two values are $+1$ and $-1$, so each weight matrix is a $\pm1$ matrix; we denote these Stage-2 weight matrices by $r_n$ and give their explicit form and structure in the next paragraph.

\paragraph{Math Hypothesis\label{sec:tpf}}
For $n\in\{2, 3, 4\}$, the $\pm1$ weight matrices $r_n$ obtained from the Stage~$2$ model take the following explicit form:

\begin{equation}
r_2 = \begin{bmatrix}1 & 1 \\ 1 & -1\end{bmatrix}
r_3 = \begin{bmatrix}1 & 1 & 1 & 1 \\ 1 & -1 & -1 & 1 \\ 1 & 1 & -1 & -1 \\ 1 & -1 & 1 & -1\end{bmatrix}
r_4 = \begin{bmatrix} 1 & 1 & 1 & 1 & 1 & 1 & 1 & 1\\
1 & -1 & -1 & 1 & 1 & -1 & -1 & 1 \\
1 & 1 & -1 & -1 & -1 & -1 & 1 & 1 \\
1 & -1 & 1 & -1 & -1 & 1 & -1 & 1 \\
1 & 1 & 1 & 1 & -1 & -1 & -1 & -1 \\
1 & -1 & -1 & 1 & -1 & 1 & 1 & -1 \\
1 & 1 & -1 & -1 & 1 & 1 & -1 & -1 \\
1 & -1 & 1 & -1 & 1 & -1 & 1 & -1\end{bmatrix}
\label{eq:matrices}
\end{equation}

However, in every row and column except the bottom row and the left column, the numbers of $+1$ and $-1$ match. This suggests that the matrices can be obtained by permuting the rows and columns of the tensor power of the very first matrix $r_2$, as is confirmed numerically on examples.

For each of these three matrices, you can make sure that their inverses obey the relation:

\begin{equation}
    r_n^{-1} = \frac{1}{2^{n-1}}r_n^T
    \label{eq:r_n_minus1}
\end{equation}

The relationship between the absolute values of the determinants for different orders $n$ is given by:

\begin{equation}
    |\det(r_n)| = |\det(r_2)|^{(n-1)\cdot2^{n-2}}
\end{equation}

Note that the dependence of the tensor degree $r_2$ is as follows:

\begin{equation}
    |\det(r_2)^{\ot k}| = |\det(r_2)|^{k\cdot2^{k-1}}
\end{equation}

That is, just as if the equality were true up to the permutation of rows and columns:

\begin{equation}
    r_n =_\sim r_2^{\ot (n-1)}
    \label{eq:r_nrecursive}
\end{equation}

where $=_\sim$ means "up to permutations". The proof of this fact is beyond the scope of this article, but numerical agreement has been verified on data up to $10$ qubits. A rigorous proof of this fact is given in our article~\cite{fedin2025mathematicalaspectsdecompositiondiagonal}.

\newpage 
    \section{Conclusion}
\label{sec:Chapter6} \index{Chapter6}

In this article, we have presented a practical technique that allows us to quickly test hypotheses and identify simple related subsystems in an unknown set of quantum circuits. This problem can also be approached from other angles. For example, when the parameter space of the final scheme is discrete, it may be useful to use a genetic algorithm (GA) whose genome encodes the matrix structure directly, with the values restricted to the set $\{-1,+1\}$. Likewise, the same workflow can be applied to subgroups whose parameter map is more complex than in the diagonal case, where the interpretable model is correspondingly harder to analyze. A more rigorous mathematical proof of the decomposition was provided in~\cite{fedin2025mathematicalaspectsdecompositiondiagonal}.

\section*{Acknowledgements}

We are grateful for very useful discussions with A. Belov, M. Fedorov, K. Gubarev, D. Vasiliev, D. Korzun, D. Khudoteplov, I. Sudakov, R. Khashaev. This work was supported by The Ministry of Economic Development of the Russian Federation (IGK 000000C313925P4C0002), agreement No139-15-2025-010.

\newpage


    \nocite{*}
    \bibliography{references}
    \newpage
    \section{Appendix}
\label{sec:Apendix} \index{Apendix}

\subsection{Proof of the correspondence of our problem to the convergence theorem \label{sec:sgd_proove}}

We are working in a Euclidean $m-$dimensional space, that is, $\mathcal{X}= \RR^m$. Our $\mathcal{A}$ mapping is linear and one-to-one, which means it does not contain noise. Here $\gamma_k$ is the step of the gradient descent.
In our work, we use the following dependence of the gradient descent step on the step number $k$:
\begin{equation}     \{\gamma_k\} = \{\dots\underbrace{1/m^\alpha,\ 1/m^\alpha,\ \dots,\ 1/m^\alpha}_{\lceil m^\beta\rceil \text{ times}}, \underbrace{1/(m+1)^\alpha,\ 1/(m+1)^\alpha,\ \dots,\ 1/(m+1)^\alpha}_{\lceil(m+1)^\beta \rceil \text{ times}}\dots\}
\label{eq:sgd_step2}
\end{equation}

where $\lceil x \rceil$ means the integer part of the number $x$ rounded up, which allows us to assume that the sum of the subsequences of the same values:

\begin{equation}
   \frac{1}{m^{\alpha-\beta}} \leqslant \sum_{i=1}^{\lceil m^\beta\rceil} \frac{1}{m^\alpha} \leqslant \frac{2}{m^{\alpha-\beta}}
    \label{eq:sgs_un_step2}
\end{equation}

\begin{enumerate}
    \item The input data does not contain noise, that is, the observations \(\mathbf{y}\) correspond exactly to the model \( \mathbf{y} = \mathcal{A} \mathbf{x} \).
    \begin{proof}
        The data does not contain noise by construction - they are generated by a theoretical dependence of a deterministic nature.
    \end{proof}    
    \item \( \mathcal{X} \)  is a Banach space that is strictly convex and smooth.
    \begin{proof}
        For $\RR^m$ this is obvious: for any $x,y \in \RR^m$ and $t\in [0,1]$ we have $xt+(1-t)y\in \RR^m$
    \end{proof}  
    \item The space allows for a dual mapping \( J : \mathcal{X} \to \mathcal{X}^* \), which is continuous and strictly monotonous.
    \begin{proof}
        If $\mathcal{X} \equiv \RR^m$ is a trivial isomorphism $\RR^{m*}\cong \RR^m$
    \end{proof} 
    \item \( \mathcal{A} \) is linear and continuous operator \( \mathcal{A} : \mathcal{X} \to \mathcal{Y} \). 
    \begin{proof}
        \(\mathcal{A}\) linear by definition, and linear operator in Euclidean spaces \(\RR\) is continious.
    \end{proof} 
    \item The operator \(\mathcal{A}\) has a closed image, which guarantees the existence of a solution with a minimum norm.

    \begin{proof}
        The image of a Euclidean space with a linear invertible map is a Euclidean space --- a closed manifold.
    \end{proof} 
    
    \item The sequence \(\{ \gamma_k \} \subset \RR_+ \) is positive and satisfies the following conditions:
        \begin{equation}     \sum_{k=1}^{\infty} \gamma_k = \infty \quad \text{(divergence of the sum of steps)}, \end{equation}
        \begin{equation}     \sum_{k=1}^{\infty} \gamma_k^2 < \infty \quad \text{(convergence of the sum of square steps)}. \end{equation}

    \begin{proof}
        
        According to our dependence of the gradient descent step on the step number, we can group the terms and obtain its convergence equivalence according to \eqref{eq:sgs_un_step}:
        
        \begin{equation}
             \sum_{k=1}^{\infty} \gamma_k \geqslant \sum_{m=1}^\infty \frac{1}{m^{\alpha-\beta}} = \infty \Rightarrow \alpha-\beta \leqslant 1
        \end{equation}
        
        whereas:
        
        \begin{equation}
            \sum_{k=1}^{\infty} \gamma_k^2 \leqslant 2 \sum_{m=1}^\infty \frac{1}{m^{2\alpha-\beta}} < \infty \Rightarrow 2\alpha-\beta > 1
        \end{equation}
        
        You can also replace \(m^\beta\) with \(Bm^\beta\) and \(1/m^\alpha\) with \(A/m^\alpha\), which will not affect convergence.

        Therefore, $\alpha-1\leqslant\beta<2\alpha-1$.
        
        \end{proof}

    \item The initial approximation of the linear mapping is chosen arbitrarily.

    \begin{proof}

        We initialize the parameters arbitrarily, according to the Section \ref{sec:linear_model}.
    
    \end{proof}
    
    \item At each iteration, a data element for SGD is randomly selected, which ensures the stochasticity of the method.
    \begin{proof}
        This is implemented by the SGD optimizer that we use.
    \end{proof}
\end{enumerate}

\subsection{Principal Component Analysis (PCA)\label{sec:PCA}}

The Principal Component Analysis (PCA) method is a classic linear data transformation technique used to reduce dimensionality in multidimensional data analysis and machine learning tasks. It allows us to find an orthogonal basic representation of the feature space in which the first few directions (the principal components) preserve as much of the variance of the original data as possible. The essence of the method is to find a linear mapping that projects vectors from the source space $\mathbb{R}^d$ into a space of smaller dimension $\mathbb{R}^k$ with minimal loss of information in the sense of sample spread.

Let's have a sample of $N$ observations $x_1,\ldots, x_N\in\mathbb{R}^d$ organized into a matrix $X\in\mathbb{R}^{N\times d}$. It is assumed that the data is pre-centered: $\sum_{i=1}^N x_i = 0$. The PCA method is based on the spectral decomposition of the covariance matrix of the sample:

\begin{equation}
\Sigma = \frac{1}{N} X^T X.
\end{equation}

The eigenvectors of this matrix, sorted in descending order of the corresponding eigenvalues, indicate the directions of the main components. Projecting the vectors $x_i$ onto the first $k$ components, we get a low-dimensional representation of the data:
\begin{equation}
z_i = W^T x_i, \quad W \in \mathbb{R}^{d \times k},    
\end{equation}

where the columns $W$ are the $k$ eigenvectors $\Sigma$ corresponding to the largest eigenvalues.

The method was first proposed by K. Pearson\cite{pearson1901}, and his formalization in terms of linear algebra was further developed in the middle of the 20th century. In the context of machine learning and statistical learning, PCA has been widely used since the $1990s$. One of the first systematic applications of PCA in machine learning in modern classification and visualization tasks is the work of \cite{jolliffe2002}, as well as early research on face recognition using PCA, in particular \cite{turk1991}, where the method was used as a basis for constructing ''eigenfaces'', named by analogy with the eigenvectors of the matrices.

\subsection{Proof of the upper bound on the number of $CNOT$s for a diagonal two-qubit operator\label{sec:threeCNOT}}

We know that two $CNOT$ gates are sufficient for a two-qubit circuit, because ~\cite{threeCNOT} was proved that any gate with $h_z=0$ in a Weyl chamber can be implemented using only two $CNOT$ gates and local unitary elements.

The diagonal operator according to Fig. \ref{fig:simple_2qubit} has the form:

\begin{equation}
\begin{aligned}
U_{\text{tail}}(\varphi_3) 
&= \Pg \otimes R_Z(\varphi_3) + \Pu \otimes X R_Z(\varphi_3) X =\\
&= \Pg \otimes R_Z(\varphi_3) + \Pu \otimes R_Z(-\varphi_3) =\\
&= \diag\left(R_Z(\varphi_3), R_Z(-\varphi_3)\right) =\\
&= \exp \left(\diag\left(-\frac{i}{2}\varphi_3, \frac{i}{2}\varphi_3, \frac{i}{2}\varphi_3, -\frac{i}{2}\varphi_3\right) \right) =\\ &= \exp(-\frac{i}{2} Z \ot Z \varphi_3) \text{ where } Z = \begin{bmatrix}
    1&0\\0&-1
\end{bmatrix}
\end{aligned}
\end{equation}

This shows that for a diagonal unitary $U\in \mathbf{SU}(4)$, the nonlocal part of the gate in the sense of the formula \eqref{eq:weylcam} corresponds to an interaction of the form $i\cdot\ln(U)= g_z Z\otimes Z$, where $g_z= -\frac{1}{2}\varphi_3$. After sorting the canonical coordinates in the Weyl chamber, we get:

\begin{equation}
(h_x, h_y, h_z) = (g_z, 0, 0) \quad \Rightarrow \quad h_z = 0.    
\end{equation}

Thus, the condition of the theorem is fulfilled.

\end{document}